\documentclass[prd,aps,a4paper,twocolumn,nofootinbib]{revtex4-2}

\usepackage[utf8]{inputenc}
\usepackage[normalem]{ulem}

\usepackage{graphicx,psfrag}
\usepackage{mathrsfs}
\usepackage{amsmath,amsfonts,amssymb}
\usepackage{multirow}
\usepackage{comment}
\usepackage{bm}
\usepackage{epstopdf}
\usepackage{color}
\usepackage{mathrsfs}
\usepackage{comment}
\usepackage{multirow}
\usepackage{hyperref}
\usepackage{diagbox} % for up-to date system 
\usepackage{amssymb,amsthm,wasysym}
\usepackage{capt-of}
\usepackage{physics}
\usepackage{calligra} \DeclareMathAlphabet{\mathcalligra}{T1}{calligra}{m}{n} \DeclareFontShape{T1}{calligra}{m}{n}{<->s*[2.2]callig15}{}

\def\mbar{\overline{m}}

\def\rmd{{\rm d}}

\newcommand{\be}{\begin{equation}}
\newcommand{\ee}{\end{equation}}

\newcommand{\RN}[1]{%
  \textup{\uppercase\expandafter{\romannumeral#1}}%
}

\definecolor{gray}{rgb}{0.5,0.5,0.5}
\definecolor{cyan}{rgb}{0,0.9,0.9}
\definecolor{orange}{rgb}{0.9,0.5,0}
\definecolor{magenta}{rgb}{1,0,1}
\definecolor{purple}{rgb}{0.8,0.4,0.8}
\definecolor{darkgreen}{rgb}{0,.8,0}
\definecolor{turquoise}{rgb}{0.25,0.88,0.82}
\definecolor{turquoise}{rgb}{0,0,0.9}

\begin{document}

\interfootnotelinepenalty=10000
\raggedbottom

\title{Asymptotic gravitational-wave fluxes from a spinning test body on generic orbits around a Kerr black hole} 

\author{Viktor Skoup\'y$^{1,\,2}$}
\email{viktor.skoupy@asu.cas.cz}
\author{Georgios Lukes-Gerakopoulos$^1$}
\author{Lisa V.\ Drummond$^3$}
\author{Scott A.\ Hughes$^3$}
\affiliation{Department of Physics and MIT Kavli Institute, MIT, Cambridge, MA 02139 USA}
\affiliation{${}^1$Astronomical Institute of the Czech Academy of Sciences, Bo\v{c}n\'{i} II 1401/1a, CZ-141 00 Prague, Czech Republic}   
\affiliation{${}^2$Institute of Theoretical Physics, Faculty of Mathematics and Physics, Charles University, CZ-180 00 Prague, Czech Republic}
\affiliation{${}^3$Department of Physics and MIT Kavli Institute, MIT, Cambridge, MA 02139 USA}

\begin{abstract}
This work provides gravitational wave energy and angular momentum asymptotic fluxes from a spinning body moving on generic orbits in a Kerr spacetime up to linear in spin approximation. To achieve this, we have developed a new frequency domain Teukolsky equation solver that calculates asymptotic amplitudes from generic orbits of spinning bodies with their spin aligned with the total orbital angular momentum. However, the energy and angular momentum fluxes from these orbits in the linear in spin approximation are appropriate for adiabatic models of extreme mass ratio inspirals even for spins non-aligned to the orbital angular momentum. To check the newly obtained fluxes, they were compared with already known frequency domain results for equatorial orbits and with results from a time domain Teukolsky equation solver called {\it Teukode} for off-equatorial orbits. The spinning body framework of our work is based on the Mathisson-Papapetrou-Dixon equations under the Tulczyjew-Dixon spin supplementary condition.
\end{abstract}

\maketitle

\section{Introduction}
\label{sec:intro}

Future space-based gravitational-wave (GW) detectors, like the Laser Interferometer Space Antenna (LISA) \cite{LISA}, TianQin \cite{TianQin}  or Taiji \cite{Taiji}, are designed to detect GWs from sources emitting in the $mHz$ bandwidth like the extreme mass ratio inspirals (EMRI). An EMRI consists of a primary supermassive black hole and a secondary compact object, like a stellar-mass black hole or a neutron star, which is orbiting in close vicinity around the primary.  Due to gravitational radiation reaction, the secondary slowly inspirals into the primary, while the EMRI system is emitting GWs to infinity. Since signals from EMRIs are expected to overlap with other systems concurrently emitting GW in the $mHz$ bandwidth \cite{LISA}, matched filtering will be employed for the detection and parameter estimation of the received GW signals. This method relies on comparison of the signal with GW waveform templates and, thus, these templates must be calculated in advance and with an accuracy of the GW phases up to fractions of radians \cite{Babak:2017}. With this level of accuracy, it is anticipated that the detection of GWs from EMRIs will provide an opportunity to probe in detail the strong gravitational field near a supermassive black hole \cite{Babak:2017}.

Several techniques have been employed to model an EMRI system and the GWs it is emitting. The backbone of these techniques is the perturbation theory \cite{Poisson:2011,Pound:2021,Barack:2019} in which the secondary body is treated as a point particle moving in a background spacetime. Such an approach is justified,  because the mass ratio $q=\mu/M$ between the mass of the secondary $\mu$ and the mass of the primary $M$ lies between $10^{-7}$ and $10^{-4}$. The particle acts as a source to a gravitational perturbation to the background spacetime and conversely the perturbation exerts a force on the particle \cite{Barack:2019}. After the expansion of the perturbation in $q$, the first-order perturbation is the source of the first-order self force and both first and second-order perturbation are sources of the second-order self force. These parts of the self-force are expected to be sufficient to reach the expected accuracy needed to model an EMRI \cite{Pound:2021}.

Another technique, which is widely used in EMRI modeling, is the 
\textit{two-timescale approximation} \cite{Hinderer:2008,Miller:2021}. This approximation relies on the separation between the orbital timescale and the inspiral timescale. In an EMRI the rate of energy loss $\dot{E}$ over the energy $E$ is $\dot{E}/E =\order{q}$, which implies that the time an inspiral lasts is $\order{q^{-1}}$. Hence, the inspiraling time is much longer than the orbital timescale $\order{q^0}$. Moreover, since the mass ratio $q$ is very small, the deviation from the trajectory, which the secondary body would follow without the self force, is very small as well. Hence, an EMRI can be modelled as a secondary body moving on an orbit in a given spacetime background with slowly changing orbital parameters; this type of modelling is called \textit{adiabatic approximation} \cite{Fujita:2020,Hughes:2021,Skoupy:2022,Isoyama:2022}. 

For a nonspinning body inspiraling into a Kerr black hole the phases of the GW can be expanded in the mass ratio \cite{Hinderer:2008} as
\begin{equation}
    \Phi_\mu(t) = \frac{1}{q} \Phi_\mu^0(qt) + \Phi_\mu^1(qt) + \order{q} \, ,
\end{equation}
where the first term on the right hand side is called adiabatic and the second postadiabatic term. The adiabatic term can be calculated from the averaged dissipative part of the first-order self force, while the postadiabatic term is calculated from several other parts of the self force. Namely, from the rest of the first-order self force, i.e., the oscillating dissipative part and the conservative part, and from the averaged dissipative part of the second-order self force \cite{Pound:2021}. To accurately model the inspiral up to radians, the postadiabatic term cannot be neglected. 

So far we have discussed the case of a nonspinning secondary body, however, to accurately calculate waveforms for an EMRI, one must also include the spin of the secondary.  To understand why, it is useful to normalize the spin magnitude of the secondary $S = \order{\mu^2}$ as $\sigma = S/(\mu M) = \order{q}$ \cite{Hartl:2003a}. For example, if the spinning body is set to be an extremal Kerr black hole, i.e. $S=\mu^2$, then $\sigma=q$. Thus, the contribution of the spin of the secondary to an EMRI evolution is of postadiabatic order.

The adiabatic term in the nonspinning case can be found from the asymptotic GW fluxes to infinity and to the horizon of the central back hole. This stems from the flux-balance laws which have been proven for the evolution of energy, angular momentum and the Carter constant for nonspinning particle in Ref.~\cite{Sago:2006}. For spinning bodies in the linear in spin approximation the flux-balance laws have been proven just for the energy and angular momentum fluxes in Refs.~\cite{Akcay:2020,Mathews:2022}. In the nonlinear in spin case the motion of a spinning body in a Kerr background is non-integrable~\cite{Hartl:2003a}, i.e. there are more degrees of freedom than constants of motion. Ref.~\cite{Witzany:2019a} has been shown that the motion of a spinning particle in a curved spacetime can be expressed by a Hamiltonian with at least $5$ degrees of freedom. Hence, since this Hamiltonian system is autonomous, i.e. the Hamiltonian itself is a constant of motion, four other constants of motion are needed to achieve integrability. In the Kerr case, there is the energy and the angular momentum along the symmetry axis for the full equations, while in the linear in spin approximation R\"{u}diger~\cite{Rudiger:1981,Rudiger:1983} found two quasiconserved constants of motion \cite{Witzany:2019}. These quasiconserved constants can be interpreted as a projection of the spin to the orbital angular momentum and a quantity similar to the Carter constant \cite{Carter:1968c}. If the evolution of these quantities could be calculated from asymptotic fluxes, then one could calculate the influence of the secondary spin on the asymptotic GW fluxes. This, in turn, would allow us to capture the influence of the secondary spin on the GW phase for generic inspirals.

Fully relativistic GW fluxes from orbits of non-spinning particles along with the evolution of the respective inspirals were first calculated in Ref~\cite{Cutler:1994} for eccentric orbits around a Schwarzschild black hole and in Ref.~\cite{Finn:2000} for circular equatorial orbits around a Kerr black hole. Fluxes from eccentric orbits in the Kerr spacetime were calculated in Refs.~\cite{Glampedakis:2002,Shibata:1994}, while the adiabatic evolution of the inspirals was presented in Ref.~\cite{Fujita:2020}. Fully generic fluxes from a nonspinning body were calculated in Ref.~\cite{Drasco:2005kz} and were employed in Ref.~\cite{Hughes:2021} to adiabatically evolve the inspirals. The spin of the secondary was included to the fluxes in Refs.~\cite{Han:2010,Harms:2016a,Harms:2016b,Lukes-Gerakopoulos:2017,Akcay:2020}  from circular orbits in a black hole spacetime and to the quasi-circular adiabatic evolution of the orbits in Refs.~\cite{Piovano:2020,Piovano:2021,Skoupy:2021a,Rahman:2023}. In Ref.~\cite{Mathews:2022}  the first-order self force was calculated for circular orbits in the Schwarzschild spacetime. Finally, the fluxes from spinning bodies on eccentric equatorial orbits around a Kerr black hole were calculated in Ref.~\cite{Skoupy:2021b} and the adiabatic evolution in linear in spin approximation was calculated in Ref.~\cite{Skoupy:2022}.

In this work, we follow the frequency-domain method to calculate generic orbits of spinning bodies around a Kerr black hole developed in Refs.~\cite{Drummond:2022a,Drummond:2022b} and use it to find asymptotic GW fluxes from these orbits in the case when the spin is aligned with the orbital angular momentum. The results are valid up to linear order in the secondary spin, since the orbits are calculated only up to this order.

The rest of our paper is organized as follows. Section~\ref{sec:spinningParticles} introduces the motion of spinning test bodies in the Kerr spacetime and describes the calculation of the linear in spin part of the motion in the frequency domain. Section~\ref{sec:GWFluxes} presents the computation of GW fluxes from the orbits calculated in Section~~\ref{sec:spinningParticles}. Section~\ref{sec:implementationResults} describes the numerical techniques we have employed to calculate the aforementioned orbits and fluxes, and it presents comparisons of the new results with previously known equatorial limit results and with time domain results for generic off-equatorial orbits. Finally, Section~\ref{sec:Concl} summarizes our work and provides an outlook for possible extensions.

In this work, we use geometrized units where $c=G=1$. Spacetime indices are denoted by Greek letters and go from 0 to 3, null-tetrad indices are denoted by lowercase Latin letters $a,b,c,\ldots$ and go from 1 to 4 and indices of the Marck tetrad are denoted by uppercase Latin letters $A,B,C,\ldots$ and go from 0 to 3. A partial derivative is denoted with a comma as $U_{\mu,\nu} = \partial_\nu U_\mu$, whereas a covariant derivative is denoted by a semicolon as $U_{\mu;\nu} = \nabla_\nu U_\mu$. The Riemann tensor is defined as $R^\mu{}_{\nu\kappa\lambda} = \Gamma^{\mu}{}_{\nu\lambda,\kappa} - \Gamma^{\mu}{}_{\nu\kappa,\lambda} + \Gamma^{\mu}{}_{\rho\kappa} \Gamma^{\rho}{}_{\nu\lambda} - \Gamma^{\mu}{}_{\rho\lambda} \Gamma^{\rho}{}_{\nu\kappa}$, and the signature of the metric is $(-,+,+,+)$. Levi-Civita tensor $\epsilon^{\alpha\beta\gamma\delta}$ is defined as $\epsilon^{0123} = 1/\sqrt{-g}$ for rational polynomial coordinates\footnote{Note that for Boyer-Lindquist (BL) coordinates the sign is opposite since the coordinate frame in BL coordinates is right-handed whereas the coordinate frame in rational polynomial coordinates is left-handed.}.

\section{Motion of a spinning test body}
\label{sec:spinningParticles}

The motion of an extended test body in the general relativity framework was first addressed by Mathisson in \cite{Mathisson:1937zz,Mathisson:2010} where he introduced the concept of a ``gravitational skeleton'', i.e., an expansion of an extended body using its multipoles. If we wish to describe the motion of a compact object, like a black hole or a neutron star, then we can restrict ourselves to the pole-dipole approximation \cite{Hartl:2003a}, where the aforementioned expansion is truncated to the dipole term and all the higher multipoles are ignored. In this way, the extended test body is reduced to a body with spin and the respective stress-energy tensor can be written as \cite{dixon1979isolated}
\begin{multline}
    T^{\mu\nu} = \int \dd \tau \Bigg( P^{(\mu} v^{\nu)} \frac{\delta^4(x^\rho-z^\rho(\tau))}{\sqrt{-g}} \\ - \nabla_\alpha \qty(S^{\alpha(\mu} v^{\nu)} \frac{\delta^4(x^\rho-z^\rho(\tau))}{\sqrt{-g}} ) \Bigg)
\end{multline}
where $\tau$ is the proper time, $P^\mu$ is the four-momentum, $v^\mu = \dv*{z^\mu}{\tau}$ is the four-velocity, $S^{\mu\nu}$ is the spin tensor and $g$ is the determinant of the metric. Note that $x^\mu$ denotes arbitrary point of the spacetime and $z^\mu(\tau)$ denotes the position of the body parameterized by the proper time. 

From the conservation law $T^{\mu\nu}{}_{;\nu} = 0$ the Mathisson-Papapetrou-Dixon (MPD) equations \cite{Mathisson:2010,Papapetrou:1951pa,Dixon:1970zza} can be derived as
\begin{subequations}
 \label{eq:MPEQs}
\begin{align} 
    \frac{{\rm D}P^\mu}{\rmd \tau} &= - \dfrac{1}{2} \; {R^\mu}_{\nu\rho\sigma} \; v^\nu \; S^{\rho\sigma} \; , \\ 
    \frac{{\rm D}S^{\mu\nu}}{\rmd \tau} & = P^\mu v^\nu - P^\nu v^\mu 
\end{align}
\end{subequations}
where $R^{\mu}{}_{\nu\rho\sigma}$ is the Riemann tensor. However, this system of equations is underdetermined because one has the freedom in choosing the centre of mass which is tracked by the solution of these equations. To close the system, a so called spin supplementary condition (SSC) must be specified. In this work we use the Tulczyjew-Dixon \cite{tulczyjew1959motion,Dixon:1970zza} SSC
\begin{align}\label{eq:SSC}
    S^{\mu\nu} P_\mu = 0  \; .
\end{align}
Under this SSC the mass of the body
\begin{equation}\label{eq:mass}
    \mu = \sqrt{-P^\mu P_\mu}
\end{equation}
and the magnitude of its spin
\begin{equation}\label{eq:spin_magnitude}
    S = \sqrt{S^{\mu\nu}S_{\mu\nu}/2}
\end{equation}
are conserved. The relation between the four-velocity and four-momentum reads \cite{Ehlers1977}
\begin{align}
 \label{eq:v_p_TUL}
 v^\mu = \frac{\textsf{m}}{\mu} \left(
          u^\mu + 
          \frac{ \frac{1}{2} s^{\mu\nu} R_{\nu\rho\kappa\lambda} u^\rho s^{\kappa\lambda}}
          {1 + \frac{1}{4} R_{\alpha\beta\gamma\delta} s^{\alpha\beta} s^{\gamma\delta} }
          \right)
\end{align}
where 
\begin{equation}
    u^\mu = \frac{P^\mu}{\mu} \; , \qquad s^{\mu\nu} = \frac{S^{\mu\nu}}{\mu}
\end{equation}
are specific momenta and $\textsf{m} = -p^\mu v_\mu$ is a mass definition with respect to $v_\mu$ which is not conserved under TD SSC. Note that having fixed the centre of mass as a reference point for the body allows us to view it as a particle. Hence, quite often the term ``spinning particle'' is used instead of ``spinning body''. 

From the spin tensor $s^{\mu\nu}$ and the specific four-momentum $u^\mu$ we can define the specific spin four-vector
\begin{align}
\label{eq:SpinVect}
 s_\mu = -\frac{1}{2} \epsilon_{\mu\nu\rho\sigma}
          \, u^\nu \, s^{\rho\sigma} \; 
\end{align}
for which the evolution equation
\begin{equation}
    \frac{{\rm D}s^\mu}{\dd \tau} = - u^\mu R^\ast_{\alpha\beta\gamma\delta} s^\alpha v^\beta u^\gamma s^\delta
\end{equation}
holds \cite{Suzuki:1997}, where the right dual of Riemann tensor has the form
\begin{equation}
    R^\ast_{\alpha\beta\gamma\delta} = \frac{1}{2} R_{\alpha\beta}{}^{\mu\nu} \epsilon_{\mu\nu\gamma\delta} \; .
\end{equation}
Note from Eq.~\eqref{eq:SpinVect} and the properties of $\epsilon_{\mu\nu\rho\sigma}$, it is clear that $s_\mu u^\mu = 0$.

In the context of an EMRI, it is convenient to define the dimensionless spin parameter
\begin{equation}
    \sigma = \frac{S}{\mu M} \; ,
\end{equation}
since one can show that $\sigma$ is of the order of the mass ratio $q=\dfrac{\mu}{M}$ \cite{Hartl:2003a}. For instance, if the small body is set to be an extremal Kerr black hole, then $S=\mu^2$ and hence $\sigma=q$.
Having established that $\sigma \lesssim q $, one sees that this parameter is very small in the context of EMRI. Since the adiabatic order is calculated from the geodesic fluxes \cite{Drasco:2005kz}, every correction to the trajectory and the fluxes of the order of $q$ influences the first postadiabatic order and higher order corrections are pushed to second postadiabatic order and further. By taking into account that the current consensus is that for the signals observed by LISA we need an accuracy in the waveforms up to the first postadiabatic order, it is reasonable to linearize the MPD equations in the secondary spin and discard all the terms of the order $\order{\sigma^2}$ and higher. Note that in Refs.~\cite{Drummond:2022a,Drummond:2022b} a different dimensionless spin parameter is used, which is defined as
\begin{equation}
    s = \frac{S}{\mu^2} \; .
\end{equation}
It is related to $\sigma$ as $s = \sigma/q$ and its magnitude is bounded by one.

After the linearization in $\sigma$ the relation~\eqref{eq:v_p_TUL} reads
\begin{equation}
    v^\mu = u^\mu + \order{s^2}
\end{equation}
and the MPD equations themselves simplify to 
\begin{subequations}
\begin{align} 
    \frac{{\rm D}u^\mu}{\rmd \tau} &= - \dfrac{1}{2} \; {R^\mu}_{\nu\rho\sigma} \; u^\nu \; s^{\rho\sigma} \; , \label{eq:linMPD}\\ 
    \frac{{\rm D}s^{\mu\nu}}{\rmd \tau} & = 0 
\end{align}
\end{subequations}
and
\begin{equation}
\label{eq:ParallelTransport}
    \frac{{\rm D}s^\mu}{\dd \tau} = 0 \; .
\end{equation}
Eq.~\eqref{eq:ParallelTransport} is the equation of parallel transport along the trajectory. After rewriting this equation using the total derivative
\begin{equation}
    \dv{s^\mu}{\tau} + \Gamma^\mu{}_{\alpha\beta} u^\alpha s^\beta = 0\, ,
\end{equation}
it can be seen that to keep the equation truncated to $\order{\sigma}$, the Christoffel symbol $\Gamma^\mu{}_{\alpha\beta}$ and the four-momentum has to be effectively taken at the geodesic limit \cite{Drummond:2022a}. Thus, the parallel transport of the spin has to take place along a geodesic.

\subsection{Spinning particles in Kerr spacetime}

In this work we treat the binary system as a spinning body moving on a Kerr background spacetime, which line element in ``rational polynomial'' coordinates \cite{Visser:2007} read
\begin{multline}
    \dd s^2 = -\qty( 1 - \frac{2 M r}{\Sigma} ) \dd t^2 - \frac{4 a M r (1-z^2)}{\Sigma} \dd t \dd \phi\, + \\ \frac{\qty(\varpi^4 - a^2 \Delta (1-z^2))(1-z^2)}{\Sigma} \dd \phi^2 + \frac{\Sigma}{\Delta} \dd r^2 + \frac{\Sigma}{1-z^2} \dd z^2
\end{multline}
where
\begin{align*}
    \Sigma &= r^2 + a^2 z^2 \; , \\
    \Delta &= r^2 - 2 M r + a^2 \; , \\
    \varpi^2 &= r^2+a^2 \; .
\end{align*}
These coordinates are derived from the Boyer-Lindquist one with $z=\cos\theta$ and are convenient for manipulations in an algebraic software such as \textit{Mathematica}.

A Kerr black hole has its outer horizon located at $r_+ = M+\sqrt{M^2-a^2}$. A Kerr spacetime is equipped with two Killing vectors $\xi_{(t)}^\mu = \delta^\mu_t$ and $\xi_{(\phi)} = \delta^\mu_\phi$, which are related respectively to the stationarity and the axisymmetry of the spacetime. Additionally for the Kerr spacetime, there is also a Killing-Yano tensor in the form
\begin{multline}
    Y_{\mu\nu} \, \dd x^\mu \wedge \dd x^\nu = a z \, \dd r \wedge \qty( \dd t - a (1-z^2) \dd \phi ) \\ + r\, \dd z \wedge \qty( a \dd t - \varpi^2 \dd \phi )\, ,
\end{multline}
from which a Killing tensor can be defined as
\begin{equation}
    K_{\mu\nu} = Y_{\mu}{}^{\kappa} Y_{\nu\kappa} \; .
\end{equation}

Thanks to these symmetries, there exist two constants of motion for the spinning particle in the Kerr background 
\begin{subequations}\label{eq:COM}
\begin{align}
    E &= - u_\mu \xi_{(t)}^\mu + \frac{1}{2} \xi^{(t)}_{\mu;\nu} s^{\mu\nu} \; , \\
    J_z &= u_\mu \xi_{(\phi)}^\mu - \frac{1}{2} \xi^{(\phi)}_{\mu;\nu} s^{\mu\nu} \; ,
\end{align}
which can be interpreted respectively as the specific total energy measured at infinity and the component of the specific total angular momentum parallel to the axis of symmetry of the Kerr black hole measured at infinity.

Apart from the aforementioned constants, there are also a couple of quasi-conserved quantities \cite{Rudiger:1981,Rudiger:1983}
\begin{align}
    C_Y &= Y_{\mu\nu} u^\mu s^\nu \; , \\
    K_R &= K_{\mu\nu} u^\mu u^\nu - 2 u^\mu s^{\rho\sigma} \qty( Y_{\mu\rho;\kappa} Y^{\kappa}{}_{\sigma} + Y_{\rho\sigma;\kappa} Y^{\kappa}{}_{\mu} ),
\end{align}
\end{subequations}
for which it holds
\begin{equation}
    \dv{K_R}{\tau} = \order{\sigma^2} \, , \quad \dv{C_Y}{\tau} = \order{\sigma^2} \; .
\end{equation}
The existence of these quasi-conserved quantities causes the motion of a spinning particle in a Kerr background to be nearly-integrable in linear order in $\sigma$ \cite{Witzany:2019}. Actually, for Schwarzschild background $(a=0)$ it has been shown that the non-integrability effects appear at $\order{\sigma^2}$ \cite{Zelenka20}. $K_R$ is analog to the geodesic Carter constant $K = K_{\mu\nu} u^\mu u^\nu = l_\mu l^\mu$ (see Appendix~\ref{app:geodesics}),where $l^\mu = Y_{\nu}{}^\mu u^\nu$ can be interpreted as the total specific (geodesic) orbital angular momentum. Because of this, $C_Y$ can be interpreted as a scalar product of the spin four-vector with the total orbital angular momentum. In other words, $C_Y$ can be seen as a projection of the spin on the total orbital angular momentum.

The four-vector $l^\mu$ was used by Marck \cite{Marck:1983} and van de Meent \cite{vandeMeent:2020} to find a solution to a parallel transport along a geodesic in the Kerr spacetime, i.e. a solution to Eq.~\eqref{eq:ParallelTransport}. The resulting $s^\mu$ can be written as
\begin{equation}\label{eq:spin_solution}
    s^\mu = M \qty( \sigma_\perp \qty( \cos\psi_p \Tilde{e}_1^\mu + \sin\psi_p \Tilde{e}_2^\mu ) + \sigma_\parallel e_3^\mu )
\end{equation}
where we introduced $\sigma_\perp$ and $\sigma_\parallel$, which is a decomposition of the spin four-vector to a perpendicular component and to a parallel one, respectively, to the total orbital angular momentum; while $\Tilde{e}_1^\mu$, $\Tilde{e}_2^\mu$ and $e_3^\mu = l^\mu/\sqrt{K}$ are the legs of the Marck tetrad \cite{vandeMeent:2020}. (Note that the zeroth leg of the tetrad is taken to be along the 4-velocity of the orbiting body: $e^\mu_0 = u^\mu$.  Because $s_\mu u^\mu = 0$, this tetrad leg does not appear in $s^\mu$.) Similarly to \cite{Drummond:2022a,Drummond:2022b} we define $e_3^\mu$ with opposite sign from that \cite{vandeMeent:2020}. The definition of $C_Y$ implies that $\sigma_\parallel = C_Y/\sqrt{K}$.

Eq.~\eqref{eq:spin_solution} describes a vector precessing around $e_3^\mu$ with precession phase $\psi_p$, which fulfils the evolution equation
\begin{equation}\label{eq:spin}
    \dv{\psi_p}{\lambda} = \sqrt{K} \qty( \frac{(r^2+a^2) E - a L_z}{K+r^2} + a \frac{L_z - a(1-z^2)E}{K-a^2 z^2} ) \, ,
\end{equation}
where $\lambda$ is the Carter-Mino time, related to proper time along the orbit by $\dd\lambda = \dd\tau/\Sigma$. An analytic solution for $\psi_p(\lambda)$ can be found in \cite{vandeMeent:2020}. The precession introduces a new frequency $\Upsilon_s$ to the system. Since the perpendicular component $\sigma_{\perp}$ is multiplied by sine and cosine of the precession phase, the contribution of this component in the linear order is purely oscillating. Therefore, the constants of motion and the frequencies depend only on the parallel component $\sigma_\parallel$ as well as the GW fluxes of energy and angular momentum in linear order in spin. Because of this, we neglect the perpendicular component and focus on a trajectory of a spinning body with spin aligned to the total orbital angular momentum.

\subsection{Linearized trajectory in frequency domain}

We follow the procedure of Refs.~\cite{Drummond:2022a,Drummond:2022b}, where the bounded orbits of a spinning particle were parameterized   in Mino-Carter time as
\begin{subequations}\label{eq:linear}
\begin{align}
    u_t &= -\hat{E} + u_t^S(\lambda) \; , \\
    u_\phi &= \hat{L}_z + u_\phi^S(\lambda) \; , \\
    r &= \frac{p}{1+e\cos(\Upsilon_r \lambda + \delta\hat{\chi}_r(\lambda) + \delta\chi_r^S(\lambda))} + \mathcalligra{r}^S(\lambda) \; , \\
    z &= \sin I \cos( \Upsilon_z \lambda + \delta\hat{\chi}_z(\lambda) + \delta\chi^S_z(\lambda) ) + \mathcalligra{z}^S(\lambda)
\end{align}
with
\begin{align}
    \Upsilon_r &= \hat{\Upsilon}_r + \Upsilon_r^S \; , \\
    \Upsilon_z &= \hat{\Upsilon}_z + \Upsilon_z^S 
\end{align}
\end{subequations}
where the hatted quantities denote geodesic quantities and quantities with index S are proportional to $\sigma$.\footnote{$\Upsilon_s$ does not need to be expanded to first order in $\sigma$ because it appears in terms proportional to $\sigma$.}

This parametrization assumes that the particle oscillates between its radial and polar turning points, but, unlike in the geodesic case, which is described in Appendix \ref{app:geodesics}, the radial turning points depend on $z$ and the polar turning points depend on $r$. This dependence is encoded in the corrections $\mathcalligra{r}^S$ and $\mathcalligra{z}^S$, respectively. $\Upsilon_r$ and $\Upsilon_z$ are the radial and polar frequency, but because of the corrections $\mathcalligra{r}^S$ and $\mathcalligra{z}^S$, the radial and polar motion has also a small contribution from a combination of all the frequencies $n\Upsilon_r + k\Upsilon_z + j\Upsilon_s$, where $n$, $k$, and $j$ are integers. This parametrization assumes that a reference geodesic is given by the parameters: semi-latus rectum $p$, eccentricity $e$ and inclination $ I$ (see Appendix \ref{app:geodesics} for their definition) and the trajectory of a spinning particle has the same turning points after averaging.

With these frequencies at hand, quantities in Eq.~\eqref{eq:linear} parametrized with respect to $\lambda$ can be expanded in the frequency domain as
\begin{equation}\label{eq:fourier}
    f(\lambda) = \sum_{n,k,j} f_{nkj} e^{-i n \Upsilon_r \lambda - i k \Upsilon_z \lambda - i j \Upsilon_s \lambda} \, .
\end{equation}
In particular, $\delta\chi_r^S$ is summed only over positive and negative $n$; $\delta\chi_z^S$ is summed only over positive and negative $k$; $k$ and $j$ cannot be simultaneously zero for $\mathcalligra{r}^S$ and $n$ and $j$ cannot be simultaneously zero for $\mathcalligra{z}^S$. In our numerical calculations we truncate the $n$ and $k$ sums at $\pm n_{\rm max}$ and $\pm k_{\rm max}$. These maxima are determined empirically from the convergence of contributions to the total flux from each mode, as well as from the mode’s numerical properties; more details are shown in Sec. \ref{sec:implementationResults}. The index $j$ is summed from $-1$ to $1$.

After introducing the phases
\begin{subequations}
\begin{align}
    w_r &= \Upsilon_r \lambda \; , \\
    w_z &= \Upsilon_z \lambda \; , \\
    w_s &= \Upsilon_s \lambda\; ,
\end{align}
\end{subequations}
we can write the inverse expression for Eq.~\eqref{eq:fourier} as
\begin{equation} \label{eq:fourierComp}
    f_{knj} = \int \frac{\dd w_r \dd w_z \dd w_s}{(2\pi)^3} f(w_r, w_z, w_s) e^{i n w_r + i k w_z + i j w_s} \; .
\end{equation}
Equations~\eqref{eq:linMPD} together with the normalization of the four-velocity $u^\mu u_\mu = -1$ are then used to find the quantities~\eqref{eq:linear} in the frequency domain.

The coordinates can then be linearized with fixed phases as $r(w_r,w_z,w_s) = \hat{r}(w_r) + r^S(w_r,w_z,w_s)$, $z(w_r,w_z,w_s) = \hat{z}(w_z) + z^S(w_r,w_z,w_s)$, where the linear in spin parts can be expressed as \cite{Drummond:2022a,Drummond:2022b}
\begin{align}
    r^S &= \frac{e p \delta\chi_r^S \sin(w_r + \delta\hat{\chi}_r)}{(1+e\cos(w_r + \delta\hat{\chi}_r))^2} + \mathcalligra{r}^S \; . \\
    z^S &= -\sin I \delta\chi_z^S \sin(w_z + \delta\hat{\chi}_z) + \mathcalligra{z}^S \; .
\end{align}

For the calculation of gravitational-wave fluxes we need also the coordinate time and azimuthal coordinate. Both can be expressed as secularly growing part plus purely oscillating part, i.e.
\begin{align}
    t &= \Gamma \lambda + \Delta t(\Upsilon_r \lambda, \Upsilon_z \lambda, \Upsilon_s \lambda) \; , \\
    \phi &= \Upsilon_\phi \lambda + \Delta \phi(\Upsilon_r \lambda, \Upsilon_z \lambda, \Upsilon_s \lambda) \, ,
\end{align}
where the oscillating parts $\Delta t$ and $\Delta \phi$ cannot be separated, unlike in the geodesic case in Eq.~\eqref{eq:geotphi} where they are broke up in a $r$ and $z$ part \cite{Fujita:2009}. These oscillating parts can be calculated from the four-velocity with respect to Carter-Mino time, $U^\mu \equiv \dd x^\mu/\dd \lambda = \Sigma u^\mu \equiv \Sigma \dd x^\mu/\dd \tau$. After integrating
\begin{equation}\label{eq:dtdlambda}
    \dv{t}{\lambda} = U^t = \sum_{n,k,j} U^t_{nkj} e^{-i n \Upsilon_r \lambda - i k \Upsilon_z \lambda - i j \Upsilon_s \lambda} \; ,
\end{equation}
the $n,k,j$-mode of $\Delta t(\lambda)$ in the frequency domain Eq.~\eqref{eq:fourier} reads
\begin{equation}
    \Delta t_{nkj} = \frac{U^t_{nkj}}{-i n \Upsilon_r - i k \Upsilon_z - j \Upsilon_s} \, ,
\end{equation}
where $U^t_{nkj}$ is the harmonic mode of the four-velocity. By linearizing in spin the above equation we obtain
\begin{multline}
    \Delta t^S_{nkj} = \frac{i U^t_{S,nkj}}{n \hat{\Upsilon}_r + k \hat{\Upsilon}_z + j \Upsilon_s} - \frac{i\hat{U}^t_{nkj} (n \Upsilon_r^S + k \Upsilon_z^S)}{(n \hat{\Upsilon}_r + k \hat{\Upsilon}_z)^2} \; .
\end{multline}
The second term is zero for $j=\pm 1$ and $\Upsilon_s^S$ is not needed, since the geodesic motion is independent of $\Upsilon_s$. The linear in spin part  of the $t$ component of the four-velocity can be expressed as
\begin{equation}\label{eq:UtS}
    U^t_S = \pdv{V^t}{r} r^S + \pdv{V^t}{z} z^S - \pdv{V^t}{E} u_t^S + \pdv{V^t}{L_z} u_\phi^S
\end{equation}
where $V^t$ is given in Eq.~\eqref{eq:Vt}. Similarly for $\Delta \phi^S$, we use $U^\phi$ to get $\Delta \phi_{nkj}$ and consequently $\Delta \phi^S_{nkj}$, in which $U^\phi_S$ is as Eq.~\eqref{eq:UtS}, but instead of $V^t$ we use $V^\phi$. 

The linear in spin parts  of $\Gamma$ and $\phi$ are respectively $U^t_{S,000}$ and $U^\phi_{S,000}$ \cite{Drummond:2022b}. The coordinate-time frequencies read
\begin{subequations}
\begin{align}
    \Omega_r &= \frac{\hat{\Upsilon}_r + \Upsilon_r^S}{\hat{\Gamma} + \Gamma^S} \; , \\
    \Omega_z &= \frac{\hat{\Upsilon}_z + \Upsilon_z^S}{\hat{\Gamma} + \Gamma^S} \; , \\
    \Omega_\phi &= \frac{\hat{\Upsilon}_\phi + \Upsilon_\phi^S}{\hat{\Gamma} + \Gamma^S} \; , \\
    \Omega_s &= \frac{\hat{\Upsilon}_s}{\hat{\Gamma} + \Gamma^S} \; .
\end{align}
\end{subequations}

\section{Gravitational-wave fluxes}
\label{sec:GWFluxes}

In this work we calculate the gravitational waves generated by a spinning particle moving on a generic orbit around a Kerr black hole using the Newman-Penrose (NP) formalism. We calculate a perturbation of the NP scalar
\begin{equation}\label{eq:NPscalar}
    \Psi_4 = -C_{\alpha\beta\gamma\delta} n^\alpha \mbar^\beta n^\gamma \mbar^\delta
\end{equation}
where $C_{\alpha\beta\gamma\delta}$ is the Weyl tensor and $n^\mu$ and $\bar{m}^\mu$ are part of the Kinnersley tetrad $(\lambda_1^\mu, \lambda_2^\mu, \lambda_3^\mu, \lambda_4^\mu) = (l^\mu, n^\mu, m^\mu, \mbar^\mu)$ defined as
\begin{subequations}
\begin{align}
    l^\mu &= \qty(\frac{r^2+a^2}{\Delta}, 1, 0, \frac{a}{\Delta}) \; , \\
    n^\mu &= \frac{1}{2\Sigma}\left( \varpi^2 , - \Delta , 0, a \right) \; , \\
    m^\mu  &= \frac{\sqrt{1-z^2}}{\sqrt{2}\bar{\zeta}} \left( ia , 0,  -1, \frac{i}{1-z^2} \right) \; , \\
    \mbar^\mu  &= \frac{\sqrt{1-z^2}}{\sqrt{2}\zeta} \left( -ia , 0, -1, -\frac{i}{1-z^2} \right)
\end{align}
\end{subequations}
with
$$\zeta=r-i a z.$$ 

From the NP scalar \eqref{eq:NPscalar} we can calculate the strain at infinity using the equation
\begin{equation} \label{eq:Psi4}
    \Psi_4(r \rightarrow \infty) = \frac{1}{2} \dv[2]{h}{t} \; ,
\end{equation}
where $h = h_+ - i h_{\cross}$ is expressed using the two polarizations of the GW. The NP scalar $\Psi_4$ can be found using Teukolsky equation \cite{Teukolsky:1973ha}
\begin{equation} \label{eq:teuk}
    {}_{-2}\mathcal{O} \, {}_{-2}\psi(t,r,\theta,\phi) = 4\pi \Sigma T \; ,
\end{equation}
where $_{-2}\psi = \zeta^4 \Psi_4$, $_{-2}\mathcal{O}$ is a second order differential operator and $T$ is the source term defined from $T^{\mu\nu}$.

We solve the Eq.~\eqref{eq:teuk} in frequency domain, where it can be decomposed as
\begin{equation} \label{eq:psi_fourier}
    {}_{-2}\psi = \sum_{l,m}^{\infty} \frac{1}{2\pi} \int_{-\infty}^{\infty} \rmd \omega\, \psi_{lm\omega}(r) {}_{-2}S_{lm}^{a\omega}(z) e^{-i\omega t + i m \phi} \; .
\end{equation}
Then, Eq.~\eqref{eq:teuk} can be separated into two ordinary differential equations, one for the radial part $\psi_{lm\omega}(r)$ and one for the angular part $_{-2}S_{lm}^{a\omega}(z)$, which is called the spin-weighted spheroidal harmonics and is normalized as
\begin{equation}
    \int_{-1}^1 \abs{ _{-2}S_{lm}^{a\omega}(z) }^2 \dd z = \frac{1}{2\pi}\, .
\end{equation}

The radial equation reads
\begin{equation}
    \mathcal{D}_{lm\omega} \psi_{lm\omega}(r) = \mathcal{T}_{lm\omega}\, ,
\end{equation}
where $\mathcal{D}_{lm\omega}$ is a second order differential operator, which depends on $r$, and $\mathcal{T}_{lm\omega}$ is the source term which we describe later. Because the source term is zero around the horizon and infinity, the function $\psi_{lm\omega}(r)$ must satisfy boundary conditions at these points for the vacuum case that read \cite{Hughes:2021}
\begin{subequations}
\begin{align}
    \psi_{lm\omega}(r) &\approx C^+_{lm\omega} r^3 e^{i\omega r^\ast} \qquad r \rightarrow \infty \, , \\
    \psi_{lm\omega}(r) &\approx C^-_{lm\omega} \Delta e^{-i k_{\mathcal{H}} r^\ast} \qquad r \rightarrow r_+ \, ,
\end{align}
\end{subequations}
where $k_{\mathcal{H}}=\omega - m a/(2M r_+)$ is the frequency at the horizon and $r^\ast = \int \varpi^2/\Delta \dd r$ is the tortoise coordinate. The amplitudes at infinity and at the horizon $C^{\pm}_{lm\omega}$ can be determined using the Green function formalism as
\begin{equation}\label{eq:Cpm}
    C^\pm_{lm\omega} = \frac{1}{W} \int_{r_+}^{\infty} \frac{R^{\mp}_{lm\omega} \mathcal{T}_{lm\omega}}{\Delta^2} \dd r \, ,
\end{equation}
where $R^\mp_{lm\omega}(r)$ are the solutions of the homogeneous radial Teukolsky equation satisfying boundary conditions at the horizon and at infinity, respectively, and $W = \qty(\qty(\partial_r R^+_{lm\omega}) R^-_{lm\omega} - R^+_{lm\omega} \partial_r R^-_{lm\omega} )/\Delta$ is the invariant Wronskian.

According to \cite{Piovano:2020}, the source term can be written as
\begin{equation}\label{eq:source_term}
    \mathcal{T}_{lm\omega} = \int \dd t \dd \theta \dd \phi \Delta^2 \sum_{ab} \mathcal{T}_{ab} e^{i\omega t -i m \phi}
\end{equation}
where $ab = nn, n\bar{m}, \bar{m}\bar{m}$ and
\begin{equation}\label{eq:source_term2}
    \mathcal{T}_{ab} = \sum_{i=0}^{I_{ab}} \pdv[i]{r}\qty( f^{(i)}_{ab} \sqrt{-g} T_{ab} )
\end{equation}
with $I_{nn} = 0$, $I_{n\bar{m}} = 1$, $I_{\bar{m}\bar{m}} = 2$. Note that the functions $f_{ab}^{(i)}$, which are defined in Appendix~\ref{app:source}, are slightly different than the definition in \cite{Piovano:2020}. The projection of the stress-energy tensor into the tetrad can be written as \cite{Tanaka:1996ht}
\begin{subequations}
\begin{equation}\label{eq:Tab}
    \sqrt{-g} T_{ab} = \int \dd \tau \qty( (A^{\rm m}_{ab} + A^{\rm d}_{ab}) \delta^4 - \partial_\rho \qty( B^\rho_{ab} \delta^4 ) )
\end{equation}
where
\begin{align}
    A^{\rm m}_{ab} &= P_{(a} v_{b)} \; , \label{eq:Amab}\\
    A^{\rm d}_{ab} &= S^{c d} v_{(b} \gamma_{a)dc} + S^{c}{}_{(a} \gamma_{b)dc} v^d \; , \label{eq:Adab}\\
    B^\rho_{ab} &= S^{\rho}{}_{(a} v_{b)} \label{eq:Brhoab}
\end{align}
\end{subequations}
and the spin coefficients are defined as
\begin{equation}
    \gamma_{adc} = \lambda_{a\mu;\rho} \lambda^\mu_d \lambda^\rho_c \; .
\end{equation}

After substituting Eqs.~\eqref{eq:source_term}, \eqref{eq:source_term2}, \eqref{eq:Tab} into Eq.~\eqref{eq:Cpm} and integrating over the delta functions, the amplitudes $C^\pm_{lm\omega}$ can be computed as 
\begin{widetext}
\begin{equation}
    C^\pm_{lm\omega} = \int_{-\infty}^{\infty} \frac{\rmd \tau}{\Sigma} e^{i\omega t(\tau) - i m \phi(\tau)} I^\pm_{lm\omega}(r(\tau), z(\tau), u_a(\tau), S_{ab}(\tau)) \, ,
\end{equation}
where $I^\pm_{lm\omega}$ is defined as
\begin{equation}\label{eq:Ipmlmomega}
    I^+_{lm\omega} = \frac{\Sigma}{W}  \sum_{ab} \sum_{i=0}^{I_{ab}} (-1)^i \qty( \qty( \qty( A^{\rm m}_{ab} + A^{\rm d}_{ab} + i \qty(\omega B^t_{ab} - m B^\phi_{ab}) ) f^{(i)}_{ab} + B^r_{ab} \pdv{f^{(i)}_{ab}}{r} + B^z_{ab} \pdv{f^{(i)}_{ab}}{z} ) \dv[i]{R^\mp_{lm\omega}}{r} + B^r_{ab} f^{(i)}_{ab} \dv[i+1]{R^\mp_{lm\omega}}{r} ) \, .
\end{equation}
Explicit expressions for $A^{\rm m}_{ab}$, $A^{\rm d}_{ab}$ and $B^\mu_{ab}$ are given in Appendix \ref{app:source}.

Following a similar procedure to \cite{Drasco:2005kz}, it can be proven that the amplitudes can be written as a sum over discrete frequencies
\begin{equation}
    C^{\pm}_{lm\omega} = \sum_{m,n,k,j} C^\pm_{lmnkj} \delta(\omega - \omega_{mnkj}) \quad \text{with} \quad \omega_{mnkj} = m \Omega_\phi + n \Omega_r + k \Omega_z + j \Omega_s \; .
\end{equation}
The partial amplitudes are given by
\begin{multline}\label{eq:Cpm_lmnkj}
    C^{\pm}_{lmnkj} = \frac{1}{(2\pi)^2 \Gamma} \int_0^{2\pi} \dd w_r \int_0^{2\pi} \dd w_z \int_0^{2\pi} \dd w_s \; I^{\pm}_{lmnkj}(w_r, w_z, w_s) \\ \times \exp(i \omega_{mnkj} \Delta t(w_r, w_z, w_s) - i m \Delta \phi(w_r, w_z, w_s) + i n w_r + i k w_z + i j w_s) \, ,
\end{multline}
where $I^\pm_{lmnkj}(w_r, w_z, w_s) = I^\pm_{lm\omega_{mnkj}}(r(w_r, w_z, w_s),z(w_r, w_z, w_s),u_a(w_r, w_z, w_s),S_{ab}(w_r, w_z, w_s))$.

\end{widetext}

The strain at infinity can be expressed from Eq.~\eqref{eq:Psi4} as
\begin{equation}
    h = -\frac{2}{r} \sum_{l,m,n,k,j} \frac{C^+_{lmnkj}}{\omega_{mnkj}^2} S_{lmnkj}(\theta) e^{-i \omega_{mnkj} u + i m \phi} \, ,
\end{equation}
where $u=t-r^\ast$ is the retarded coordinate and $S_{lmnkj}(\theta) = {}_{-2}S_{lm}^{a\omega_{mnkj}}(\theta)$.

From the strain $h$ and the stress energy tensor of a GW, the averaged energy and angular momentum fluxes can be derived as
\begin{subequations}\label{eq:fluxes}
\begin{align}
    \left\langle\mathcal{F}^{E}\right\rangle &= \sum_{l,m,n,k,j} \mathcal{F}^E_{lmnkj} \; , \\
    \left\langle\mathcal{F}^{J_z}\right\rangle &= \sum_{l,m,n,k,j} \mathcal{F}^{J_z}_{lmnkj} 
\end{align}
with
\begin{align}
    \mathcal{F}^{E}_{lmnkj} &= \frac{\abs{C^+_{lmnkj}}^2 + \alpha_{lmnkj}\abs{C^-_{lmnkj}}^2}{4\pi \omega_{mnkj}^2} \; , \\
    \mathcal{F}^{J_z}_{lmnkj} &= \frac{m\qty(\abs{C^+_{lmnkj}}^2 + \alpha_{lmnkj}\abs{C^-_{lmnkj}}^2 )}{4\pi \omega_{mnkj}^3} \; ,
\end{align}
\end{subequations}
where
\begin{equation}
    \alpha_{lmnkj} = \frac{256(2Mr_+)^5 k_{\mathcal{H}} (k_{\mathcal{H}}^2+4\epsilon^2) (k_{\mathcal{H}}^2+16\epsilon^2)\omega_{mnkj}^3}{\left| \mathscr{C}_{lm\omega_{mnkj}} \right|^2} \; ,
\end{equation}
$\epsilon = \sqrt{M^2-a^2}/(4Mr_+)$, and the Teukolsky-Starobinsky constant is
\begin{align}
    \left| \mathscr{C}_{lm\omega} \right|^2 =& \left( \left( \lambda_{lm\omega} + 2 \right)^2 + 4a\omega (m - a\omega) \right) \nonumber\\ & \times \left( \lambda_{lm\omega}^2 + 36a\omega (m - a\omega) \right) \nonumber\\ & - \left( 2\lambda_{lm\omega} + 3 \right) \left( 48a\omega (m - 2a\omega) \right) \nonumber\\ & + 144 \omega^2 \left( M^2-a^2 \right)\; .
\end{align}

Since all the terms proportional to the perpendicular component $\sigma_\perp$ are purely oscillating with frequency $\Omega_s$, the only contribution to the fluxes from $\sigma_\perp$ comes from the modes with $j=\pm 1$. The amplitudes $C^\pm_{lmnkj}$ for $j=\pm 1$ are proportional to $\sigma_\perp$ and, therefore, the fluxes for $j=\pm 1$ are quadratic in $\sigma_\perp$. We can neglect them in the linear order in $\sigma$ and sum over $l$, $m$, $n$ and $k$ with $j=0$. In this work we focus on the contribution of the parallel component $\sigma_\parallel$ to the fluxes and, therefore, calculate only the $j=0$ modes. For simplicity, we omit in the rest of the article the $j$ index and write $\omega_{mnk}$, $\mathcal{F}_{lmnk}$.

Note that since the trajectory is computed up to linear order in $\sigma$, the amplitudes or the fluxes are valid up to $\order{\sigma}$ as well.

\section{Numerical implementation and results}
\label{sec:implementationResults}

In this section we describe the process of numerically calculating the orbit and the fluxes described in the previous sections. If not stated otherwise, all calculations were done in {\it Mathematica}. In some parts of these calculations we used the {\it Black Hole Perturbation Toolkit} (BHPT) \cite{BHPToolkit}. 

\subsection{Calculating the trajectory}

Our approach to calculate the linear in spin parts of the trajectory is the same as the approach described in \cite{Drummond:2022a,Drummond:2022b}. We managed to simplify the equations given in the latter papers and the respective details are given in Appendix~\ref{app:trajectory}. To calculate the geodesic motion we employed the \textit{KerrGeodesics} package of the BHPT. 

Using the aforementioned simplifications, we first calculated $u_{t,nk}^S$ and $u_{\phi,nk}^S$ as
\begin{equation}
    u_{t,nk}^S = \frac{i \mathcal{R}_{t,nk}}{n \hat{\Upsilon}_r + k \hat{\Upsilon}_z} \; , \quad u_{\phi,nk}^S = \frac{i \mathcal{R}_{\phi,nk}}{n \hat{\Upsilon}_r + k \hat{\Upsilon}_z}
\end{equation}
for $n \neq 0$ or $k \neq 0$, where $\mathcal{R}_{t,nk}$ and $\mathcal{R}_{\phi,nk}$ are Fourier coefficients of functions given in Eqs.~\eqref{eq:traj_source}.  Then, the Fourier coefficients $u_{t,00}^S$, $u_{\phi,00}^S$, $\delta\chi_{r,n}^S$, $\delta\chi_{z,k}^S$, $\mathcalligra{r}_{nk}^S$, $\mathcalligra{z}_{nk}^S$ and the frequencies' components $\Upsilon_r^S$ and $\Upsilon_z^S$ were calculated as the least squares solution to the system of linear equations \cite{Drummond:2022a}
\begin{align} \label{eq:linsys}
    \mathbf{M} \cdot \mathbf{v} + \mathbf{c} = 0\; .
\end{align}
In the system of equations~\eqref{eq:linsys}, the column vector $\mathbf{v}$ contains the unknown coefficients, the column vector $\mathbf{c}$ is given from Fourier expansion components of the functions $\mathcal{J}$, $\mathcal{V}$ and $\mathcal{P}$ in Eqs.~\eqref{eq:traj_source} that are not coefficients of the unknown quantities, while the elements of the matrix $\mathbf{M}$ are calculated from the Fourier coefficients of functions $\mathcal{F}_{r,\mathcalligra{r}}$, $\mathcal{G}_{r,\mathcalligra{r},\theta,\mathcalligra{z}}$, $\mathcal{H}_{r,\mathcalligra{r},\theta,\mathcalligra{z}}$, $\mathcal{I}_{1r,1\theta,2,3}$, $\mathcal{Q}_{\theta,\mathcalligra{z}}$, $\mathcal{S}_{r,\mathcalligra{r},\theta,\mathcalligra{z}}$, $\mathcal{T}_{r,\mathcalligra{r},\theta,\mathcalligra{z}}$, $\mathcal{U}_{1r,1\theta,2,3}$, $\mathcal{K}_{r,\mathcalligra{r},\theta,\mathcalligra{z}}$, $\mathcal{M}_{r,\mathcalligra{r},\theta,\mathcalligra{z}}$, $\mathcal{N}_{1r,1\theta}$, which are functions of the geodesic quantities and they are given in the supplemental material of \cite{Drummond:2022a}. 

In particular, the Fourier coefficients are calculated as, e.g.,
\begin{equation}
    \mathcal{R}_{t,nk} = \sum_{a,b} \mathcal{R}_t(\hat{r}(w_r^a), \hat{z}(w_z^b)) F^{a}_{n} G^{b}_k
\end{equation}
where $F^a_n$ and $G^b_k$ are matrices of a discrete Fourier transform
\begin{subequations}
\begin{align}
    F^a_n &= \exp(\frac{\pi i n}{N_r} \qty( 1 + 2a ) ) \frac{1}{N_r} \; , \\
    G^b_k &= \exp(\frac{\pi i k}{N_z} \qty( 1 + 2b ) ) \frac{1}{N_z}
\end{align}
\end{subequations}
and $N_r$ ($N_z$) is the number of points along $w_r$ ($w_z$). Each function $\mathcal{R}_t$ is evaluated at equidistant points along $w_r$ and $w_z$ as 
\begin{subequations}
\begin{align}
    w_r^a &= \frac{2\pi}{N_r} \qty( \frac{1}{2} + a) \; , \\
    w_z^b &= \frac{2\pi}{N_z} \qty( \frac{1}{2} + b)
\end{align}
\end{subequations}
where $a = 0, 1, \ldots, N_r - 1$, $b = 0, 1, \ldots, N_z - 1$. The numbers of steps along $w_r$ and $w_z$ were chosen according to the orbital parameters, i.e., a higher number of steps is needed for higher eccentricity and higher inclination. 

\begin{figure}[tb!]
    \centering
    \includegraphics{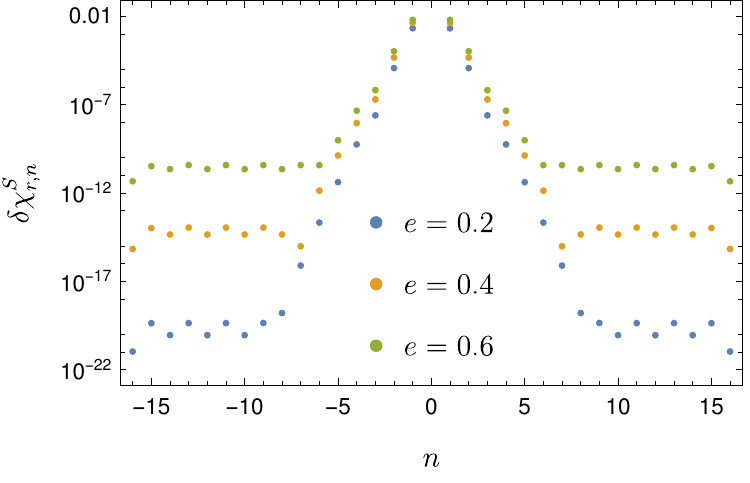}
    \caption{Fourier coefficients $\delta \chi^S_{r,n}$ for generic orbits with $a=0.9M$, $p=15$, $I=15^\circ$ and different eccentricities. Because the Fourier series is truncated at $n_{\rm max}=16$ and the coefficients have been calculated approximately using least squares, the convergence stops at certain $\pm n$.}
    \label{fig:fourier_coefficients}
\end{figure}

Actually, not all of the Fourier coefficients can be calculated accurately enough for highly eccentric and inclined orbits, as can be seen in Fig.~\ref{fig:fourier_coefficients}, where the coefficients $\delta\chi^S_{r,n}$ are plotted for different eccentricities. Fig.~\ref{fig:fourier_coefficients} shows that after a certain value of $n$ the coefficients stop decreasing. This is caused by the truncation of the series and by the fact that the system of equations is solved approximately using least squares. Similar behavior occurs for $\delta\chi^S_{z,k}$ and other Fourier series.

\subsection{Gravitational-wave fluxes}

After calculating the orbit, the partial amplitudes $C^\pm_{lmnk}$ are evaluated by numerically calculating the two-dimensional integral~\eqref{eq:Cpm_lmnkj}. The integral in Eq.~\eqref{eq:Cpm_lmnkj} is computed over one period of $w_r$ and of $w_z$; hence, we employ the midpoint rule, since the convergence is exponential  \cite{Hopper:2015}. The number of steps for the integration has been chosen as follows. We assume that the main oscillating part of the integrand comes from the exponential term. The number of oscillations in $w_{r}$ and $w_z$ is respectively $n$ and $k$. However, because of $\Delta t$ and $\Delta\phi$, the ``frequency'' of the oscillations can be higher at the turning points as can be seen in Fig.~3 in \cite{Skoupy:2021b}. In order to have enough steps in each oscillation, the number of steps in $w_r$ is calculated from the frequency of the oscillations at the pericentre ($w_r=0$) and apocentre ($w_r=\pi$) as
\begin{equation}
    \max\{\abs{16\lceil\varphi'_r(0) + n\rceil},\abs{16\lceil\varphi'_r(\pi) + n\rceil}, 32\} \, .
\end{equation}
Similarly, the number of steps in $w_z$ comes from the frequency at the turning point ($w_z=0,\pi$) and the equatorial plane ($w_z=\pi/2$) as 
\begin{equation}
    \max\{\abs{8\lceil\varphi'_z(0) + k\rceil},\abs{8\lceil\varphi'_z(\pi/2) + k\rceil}, 32\} \, ,
\end{equation}
where $\varphi_y(w_y) = \omega_{mnk}\Delta \hat{t}_y(w_y)-m\Delta\hat{\phi}_y(w_y)$, $y=r,z$. The integration over $w_s$ is trivial for $j=0$, since the function is independent of $w_s$.

The homogeneous radial Teukolsky equation solutions $R^\pm_{lmn\omega}$ have been calculated using the \textit{Teukolsky} package of the BHPT. There the radial Teukolsky equation is numerically integrated in hyperboloidal coordinates \cite{Macedo:2022} and the initial conditions are calculated by using the Mano-Sasaki-Takasugi method \cite{Mano:1996}. On the other hand, the spin-weighted spheroidal harmonics $_{-2}S_{lm}^{a\omega}$ have been calculated using the \textit{SpinWeightedSpheroidalHarmonics} package of the BHPT where the Leaver's method \cite{Leaver:1986} is employed.

Similarly as in \cite{Drasco:2005kz}, we use the symmetries of the motion to reduce the integral~\eqref{eq:Cpm_lmnkj} into a sum of four integrals over $0<w_r<\pi$, $0<w_z<\pi$. Apart from the geodesic symmetries $\hat{y}(w_y) = \hat{y}(2\pi-w_y)$, $\Delta \hat{x}_y(w_y) = -\Delta \hat{x}_y(2\pi - w_y)$, and $U^y(w_y) = -U^y(2\pi-w_y)$, where $x=t,\phi$, $y=r,z$, we used also symmetries of the linear in spin parts, which read $f(w_r,w_z) = f(2\pi - w_r, 2\pi - w_z)$ for $r^S$ and $z^S$ and $f(w_r,w_z) = -f(2\pi - w_r, 2\pi - w_z)$ for $U^r_S$, $U^z_S$, $\Delta t^S$, and $\Delta \phi^S$. Thanks to the reflection symmetry around the equatorial plane, there is also a symmetry $f(w_r,w_z) = f(w_r,w_z+\pi)$ for $r^S$, $U^r_S$, $\Delta t^S$, and $\Delta \phi^S$ and $f(w_r,w_z) = -f(w_r,w_z+\pi)$ for $z^S$ and $U^z_S$. Combining these symmetries, it is sufficient to evaluate the linear in spin parts only for $0<w_r<\pi$, $0<w_z<\pi$, which reduces the computational costs, since the evaluation of the Fourier series~\eqref{eq:fourier} is slow. After these optimizations, calculating one mode takes seconds for low eccentricities, inclinations and mode numbers, while it takes tens of seconds for high eccentricities, inclinations and mode numbers.

To extract the linear in spin part of the partial amplitudes or fluxes, i.e. their derivative with respect to $\sigma$, we use the fourth-order finite difference formula 
\begin{equation}
    f^S = \frac{ \frac{1}{12} f(-2\sigma) - \frac{2}{3} f(-\sigma) + \frac{2}{3} f(\sigma) - \frac{1}{12} f(2\sigma) }{\sigma} \, ,
\end{equation}
where $f = C^\pm_{lmnk}$, $\mathcal{F}^E$ or $\mathcal{F}^{J_z}$ and $\sigma=0.5$ in our calculations. This is necessary for comparisons with other results, since the $\order{\sigma^2}$ part of the fluxes is invalid due to the trajectory being linearized in spin.

\begin{figure}[tb!]
    \centering
    \includegraphics[width=0.45\textwidth]{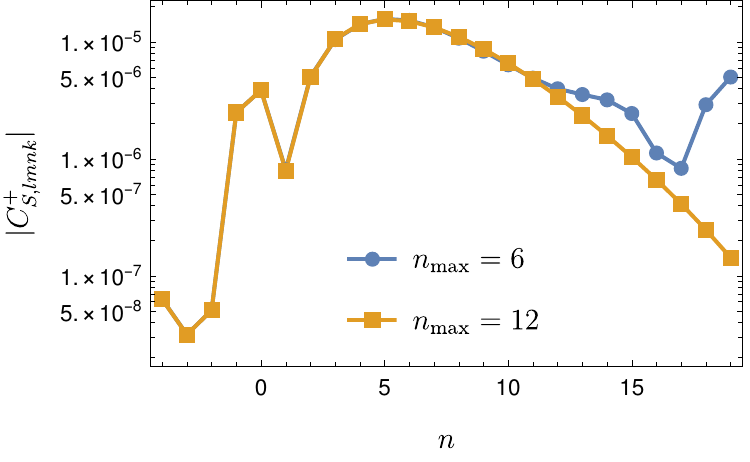}
    \includegraphics[width=0.45\textwidth]{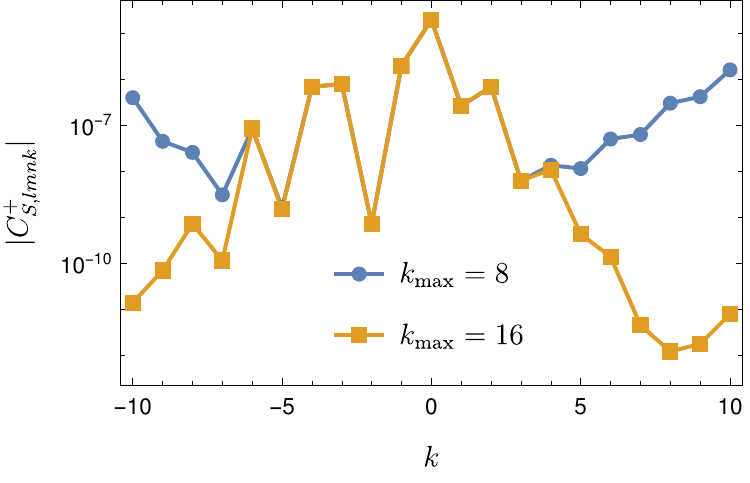}
    \caption{Top: Dependence of the linear in spin parts of the partial amplitudes for $k=0$ and different $n_{\rm max}$ for an orbit with $a=0.9M$, $p=15$, $e=0.5$, $I=15^\circ$. Bottom: Dependence of the linear in spin parts of the partial amplitudes on $k$ for $n=0$ different $k_{\rm \max}$ for an orbit with $a=0.9M$, $p=12$, $e=0.2$, $ I=60^\circ$.}
    \label{fig:truncation}
\end{figure}

Because the Fourier series~\eqref{eq:fourier} of the linear in spin part of the trajectory is truncated at $\pm n_{\rm max}$ and $\pm k_{\rm max}$, only a finite number of $n$ and $k$ modes of the amplitudes $C^\pm_{lmnk}$ and of the fluxes can be calculated accurately. In Fig.~\ref{fig:truncation} we show the dependence of the absolute value of the linear in spin parts of the amplitudes $\abs{C^{+}_{S,lmnk}}$ on $n$ and $k$ for different $n_{\rm max}$ and $k_{\rm max}$. The top panel shows amplitudes for an orbit with high eccentricity ($e=0.5$). If the Fourier series in $n$ is truncated at lower $n_{\rm max}$, the amplitudes stop being accurate after a certain value of $n$. Similarly, for an orbit with higher inclination ($I=60^\circ$) shown in the bottom panel of Fig.~\ref{fig:truncation}, when the series is truncated at lower $k_{\rm max}$, the amplitudes stop converging with $k$. Such issues have been already reported for geodesic fluxes in \cite{Kerachian:2023}.

\subsection{Comparison with the equatorial limit}

\begin{figure}[tb!]
    \centering
    \includegraphics[width=0.45\textwidth]{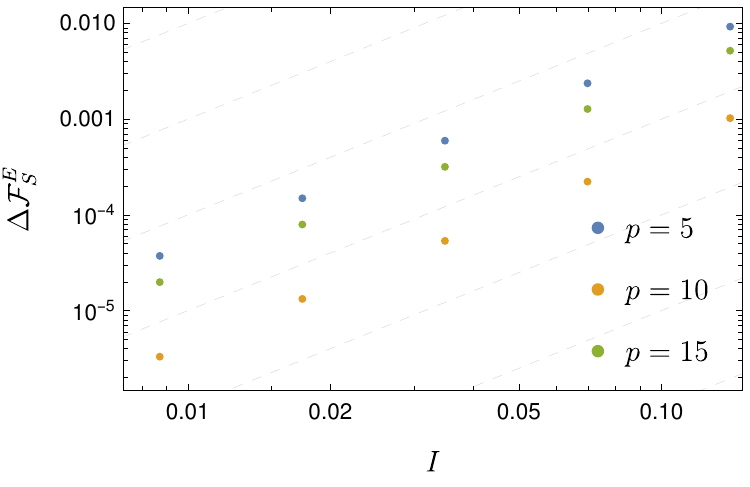}
    \caption{Relative differences of the linear in spin part of the total energy flux $\mathcal{F}^E_S$ between equatorial and nearly equatorial  cases of nearly spherical orbits for different semi-latus rectum $p$. The dashed gray lines indicate the $\order{I^2}$ behavior.}
    \label{fig:comparison_spherical}
\end{figure}

To verify our results with the equatorial limit ($I \rightarrow 0$), we have compared the frequency domain results for several inclinations with a frequency domain code for equatorial orbits \cite{Skoupy:2021b}. First, we have calculated the sum of the total energy flux over $l$ and $m$ for nearly spherical orbits with inclinations $I=0.5^\circ,1^\circ,2^\circ,4^\circ,8^\circ$. We plot the relative difference $\Delta \mathcal{F}^E_S= \abs{1-\mathcal{F}^E_S/\mathcal{F}^E_{S,I=0}}$ against $I$ in logarithmic scale in both axes in Fig.~\ref{fig:comparison_spherical}. This way, we have verified that the linear in spin part $\mathcal{F}^E_S$ asymptotically approaches the equatorial limit as $I\rightarrow 0$ with an $\order{I^2}$ difference convergence.

\begin{figure}[tb!]
    \centering
    \includegraphics[width=0.45\textwidth]{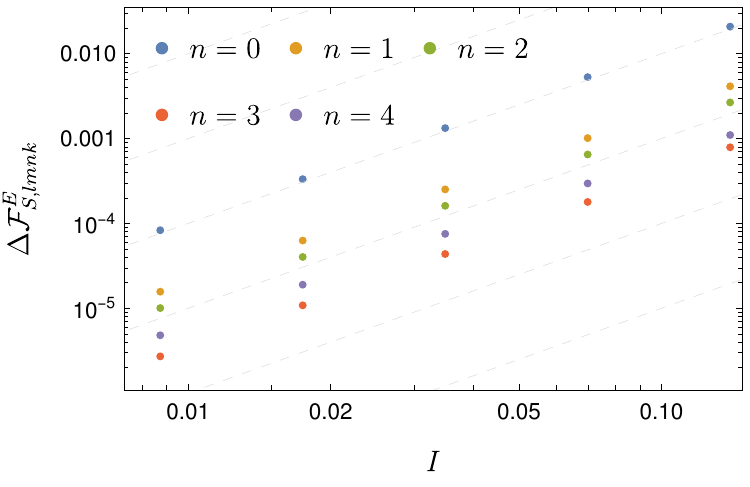}
    \includegraphics[width=0.45\textwidth]{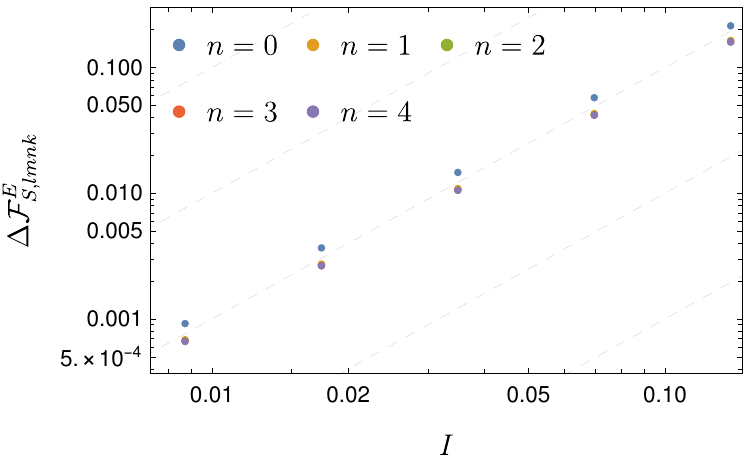}
    \caption{Relative differences of the linear in spin part of the total energy flux $\mathcal{F}^E_{S,lmn0}$ between equatorial and nearly equatorial eccentric orbits with $a=0.9M$, $p=12$, $e=0.3$. The top panel shows modes with $l=2$, $m=2$ and the bottom panel shows $l=5$, $m=4$. The dashed gray lines show the $\order{I^2}$ behavior.}
    \label{fig:comparison_generic}
\end{figure}

Similar procedure has been repeated for the eccentric orbits. We have computed the $l$, $m$, $n$ with $k=0$ modes of the energy flux $\mathcal{F}^E_{S,lmnk}$ for different inclinations $I$ and plot the relative differences $\Delta \mathcal{F}^E_{S,lmnk} = \abs{1-\mathcal{F}^E_{S,lmnk}/\mathcal{F}^E_{S,lmnk,I=0}}$ in Fig.~\ref{fig:comparison_generic}. We again see that for all the modes the relative difference in fluxes $\mathcal{F}^E_{S,lmnk}$ follows an $\order{I^2}$ convergence as $I\rightarrow 0$. This behavior agrees with the behavior of a Post-Newtonian expansion of nearly-equatorial geodesic fluxes in Refs.~\cite{Sago:2006,Sago:2015}, because the parameters $y$ and $Y$ in these references are $\order{I^2}$.

\subsection{Comparison of frequency and time domain results}

To further verify the frequency domain calculation of the fluxes $\mathcal{F}^{E}$ and $\mathcal{F}^{J_z}$, we compared them with fluxes calculated using time domain Teukolsky equation solver \textit{Teukode} \cite{Harms:2014}. This code solves the (2+1)-dimensional Teukolsky equation with spinning-particle source term in hyperboloidal horizon-penetrating coordinates. The fluxes of energy and angular momentum are extracted at the future null infinity. The numerical scheme consists of a method of lines with sixth order finite difference formulas in space and fourth order Runge-Kutta scheme in time. 

\begin{figure}[tb!]
    \centering
    \includegraphics[width=0.46\textwidth]{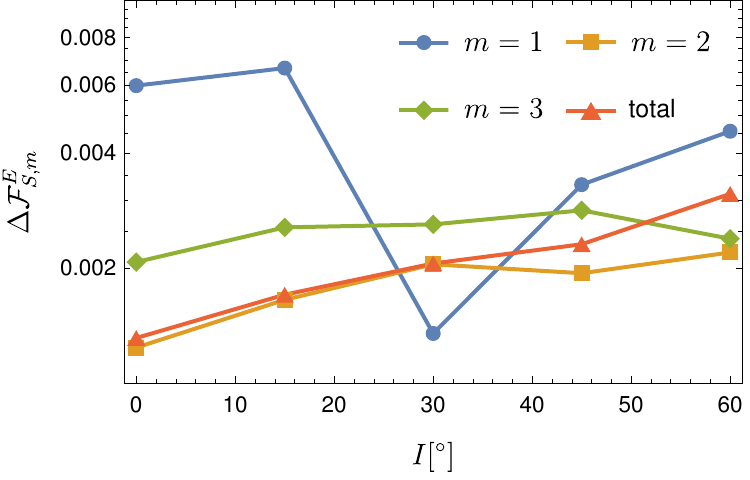}
    \includegraphics[width=0.46\textwidth]{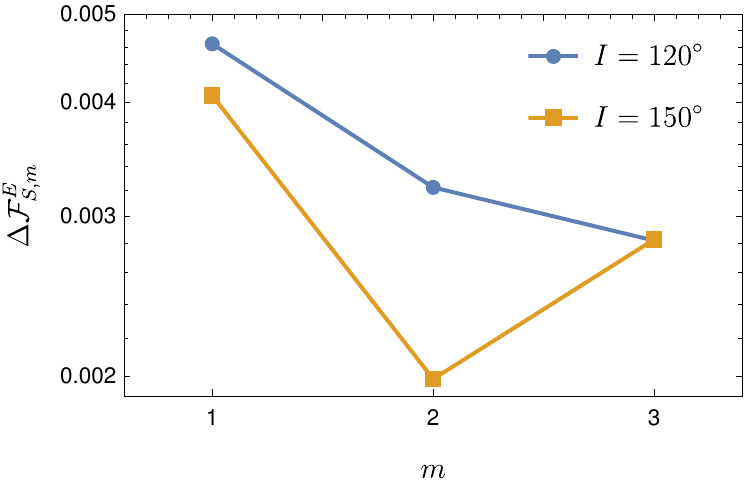}
    \caption{Relative differences of the linear in spin part of the fluxes $\mathcal{F}^{E}_{S,m}$ between time domain and frequency domain calculations for different inclinations and $m$ for nearly spherical orbits with $a=0.9M$ and $p=10$. The top panel shows prograde orbits and the bottom panel shows retrograde orbits.}
    \label{fig:error_spherical}
\end{figure}

First, we compare the computation of energy fluxes to infinity from nearly spherical orbits, i.e. orbits with $e=0$. For details about the time domain calculation of the trajectory and the fluxes see Appendix~\ref{app:time_domain}. Since the time domain outputs $m$-modes of the flux, we summed the frequency domain flux over $l$ and $k$ (for spherical orbits, only the $n=0$ modes are nonzero). In Fig.~\ref{fig:error_spherical}, we show the relative difference between the time-domain and frequency-domain-computed linear in spin part of the energy flux $\Delta \mathcal{F}^E_{S,m} = \abs{1-\mathcal{F}^{E,{\rm fd}}_{S,m}/\mathcal{F}^{E,{\rm td}}_{S,m}}$ for several inclinations $I$ and azimuthal numbers $m$. The top panel shows the dependence of the relative difference on $I$ for prograde orbits and the lower panel shows the dependence on $m$ for retrograde orbits. We can see that the error is at most $6\times 10^{-3}$ which is around the reported accuracy of Teukode in our previous paper \cite{Skoupy:2021b}. The error of the frequency domain comes from the truncation of the Fourier expansion to $n_{\rm max}$ and $k_{\rm \max}$ and from the summation of the fluxes over $l$ and $k$. On top of that, one has to take into account that the relative error of linearization of both the time domain and frequency domain flux using the fourth-order finite difference formula is around $10^{-5}$.

\begin{table}[tb!]
    \centering
    \begin{tabular}{c|c|c|c|c|c}
$p$ & $e$ & $I/^\circ$ & $m$ & $\mathcal{F}^E_{S,m}$ & $\Delta \mathcal{F}^E_{S,m}$ \\ \hline
 $10$ & $0.1$ & $15$ & $2$ & $-2.8259 \times 10^{-6}$ & $1\times 10^{-3}$ \\
 $12$ & $0.2$ & $30$ & $1$ & $-1.1954 \times 10^{-7}$ & $2\times 10^{-5}$ \\
 $12$ & $0.2$ & $30$ & $2$ & $-1.0488 \times 10^{-6}$ & $1\times 10^{-3}$ \\
 $12$ & $0.2$ & $30$ & $3$ & $-1.4210 \times 10^{-7}$ & $3\times 10^{-3}$ \\
 $12$ & $0.2$ & $60$ & $2$ & $-8.0550 \times 10^{-7}$ & $5\times 10^{-4}$ \\
 $15$ & $0.5$ & $15$ & $2$ & $-4.2936 \times 10^{-7}$ & $2\times 10^{-3}$ \\
    \end{tabular}
    \caption{Relative differences $\Delta \mathcal{F}^E_{S,m}$ of the linear in spin part of the energy flux $\mathcal{F}^E_{S,m}$ between frequency domain and time domain computations for given orbital parameters and azimuthal number $m$. All orbits have $a=0.9M$.}
    \label{tab:error_generic}
\end{table}

Next we moved to generic orbits. We have summed the energy flux over $l$, $n$ and $k$ for given $m$ and orbital parameters, in order to calculate the relative difference between the linear part of frequency domain fluxes and time domain fluxes $\Delta \mathcal{F}^E_{S,m}$. The results are presented in Table~\ref{tab:error_generic}. In this case, the relative difference is at most $3\times 10^{-3}$. 

Appendix~\ref{app:tables} shows plots of linear in spin calculations of the amplitudes and of the fluxes and some reference data tables.

\section{Summary} \label{sec:Concl}

In this work we provided asymptotic GW fluxes from off-equatorial orbits of spinning bodies in the Kerr spacetime. In our framework the spin of the small body is parallel to the orbital angular momentum and the calculations are valid up to linear order in the spin.

We employed the frequency-domain calculation of the orbits of spinning particles which was introduced in \cite{Drummond:2022a,Drummond:2022b}. In this setup, the linear in spin part of the trajectory is solved in the frequency domain using MPD equations under TD SSC. We extended this setup to calculate the corrections to the coordinate time $\Delta t^S$ and the azimuthal coordinate $\Delta \phi^S$.

We calculated GW fluxes from the aforementioned orbits using the Teukolsky equation. To do that, we constructed the source of the Teukolsky equation for off-equatorial orbits of spinning particles for spin parallel to the orbital angular momentum. Then, by using this source, we developed a new frequency-domain inhomogeneous Teukolsky equation solver in \textit{Mathematica}, which delivers the GW amplitudes $C^\pm_{lmnk}$ at infinity and at the horizon. Having these amplitudes allowed us to calculate the total energy and angular momentum fluxes, whose validity is up to linear order in the spin. Since at the linear order in spin the fluxes are independent of the precessing perpendicular component of the spin, our approach to compute the fluxes is sufficient for any linear in spin configuration.

We numerically linearized the fluxes and compared the results for nearly equatorial orbits with previously known frequency domain results \cite{Skoupy:2021b} for equatorial orbits to verify their validity in the equatorial limit. We found that the difference of the off-equatorial and equatorial flux behaves as $\order{I^2}$. Furthermore, we compared the off-equatorial results with time domain results obtained by time domain Teukolsky equation solver \textit{Teukode}. For different orbital parameters and azimuthal numbers $m$ the relative difference is around $10^{-3}$, which is the current accuracy of computations produced by \textit{Teukode}.

This work is a part of an ongoing effort to find the postadiabatic terms \cite{vandeMeent:2018a,Hughes:2021,Wardell:2021,Lynch:2022,Mathews:2022,Skoupy:2022} needed to model EMRI waveforms accurately enough for future space-based gravitational wave observatories like LISA. Our work can be extended to model adiabatic inspirals of a spinning body on generic orbits in a Kerr background as we have done for the equatorial plane case in Ref.~\cite{Skoupy:2022}; however, to achieve this the flux of the Carter-like constants $K_R$ and the parallel component of the spin $C_Y$ must be derived first. In the near future, the new frequency-domain Teukolsky equation solver \textit{Mathematica} code is planned to be published in the {\it Black Hole Perturbation Toolkit} repository.

%----------------------------------------

\begin{acknowledgments}
VS and GLG have been supported by the fellowship Lumina Quaeruntur No.\ LQ100032102 of the Czech Academy of Sciences. V.S. acknowledges support by the project ``Grant schemes at CU'' (reg.no. CZ.02.2.69/0.0/0.0/19\_073/0016935). We would like to thank Vojt\v{e}ch Witzany and Josh Mathews for useful discussions and comments. This work makes use of the Black Hole Perturbation Toolkit. Computational resources were provided by the e-INFRA CZ project (ID:90140), supported by the Ministry of Education, Youth and Sports of the Czech Republic.  LVD and SAH were supported by NASA ATP Grant 80NSSC18K1091, and NSF Grant PHY-2110384.
\end{acknowledgments}

\appendix

\section{Geodesic motion in Kerr}
\label{app:geodesics}

In this Appendix we briefly discuss aspects of geodesic motion in the Kerr spacetime.

The specific energy
\begin{align}
    E &= -u_t
\end{align}
and the specific angular momentum along the symmetry axis 
\begin{align}
    L_z &= u_\phi
\end{align}
are conserved thanks to two respective Killing vectors.
Carter in Ref.~\cite{Carter:1968c} found a third constant
\begin{equation}
 K = K_{\mu\nu} u^\mu u^\nu \, ,
\end{equation}
and formulated the equations of motion as
\begin{subequations}
\begin{align}
    \dv{t}{\lambda} &= V_t(r,z,E,L_z) \, , \\
    \dv{r}{\lambda} &= \pm \sqrt{R(r,E,L_z,K)} \, , \\
    \dv{z}{\lambda} &= \pm \sqrt{Z(z,E,L_z,K)} \, , \\
    \dv{\phi}{\lambda} &= V_\phi(r,z,E,L_z) \, ,
\end{align}
\end{subequations}
where
\begin{subequations}
\begin{align}
    V^t &= \frac{r^2+a^2}{\Delta} \qty((r^2+a^2)E - a L_z) - a^2 E (1-z^2) + a L_z ,\label{eq:Vt} \\
    R &= \qty((r^2+a^2)E - a L_z)^2 - \Delta \qty( K + r^2) \, , \\
    Z &= -\qty( (1-z^2) a E - L_z )^2 + ( 1 - z^2) \qty( K - a^2 z^2 ) \, , \\
    V^\phi &= \frac{a}{\Delta} \qty((r^2+a^2)E - a L_z) + \frac{L_z}{1-z^2} - a E \, ,
\end{align}
\end{subequations}
These equations are parameterized with Carter-Mino time $\dv*{\tau}{\lambda} = \Sigma$. The motion in $r$ oscillates between its radial turning points $r_1$ and $r_2$ with  frequency $\Upsilon_r$ and, similarly, the $z$-motion oscillates between its polar turning points $\pm z_1$ with frequency $\Upsilon_z$. Moreover, the evolution of $t$ and $\phi$ can be written as
\begin{subequations}\label{eq:geotphi}
\begin{align}
    t(\lambda) &= \Gamma \lambda + \Delta t_r(\lambda) + \Delta t_z(\lambda) \, , \\
    \phi(\lambda) &= \Upsilon_\phi \lambda + \Delta \phi_r(\lambda) + \Delta \phi_z(\lambda) \, ,
\end{align}
\end{subequations}
where $\Gamma$ and $\Upsilon_\phi$ are average rates of change of $t$ and $\phi$; while $\Delta t_r$ with $\Delta\phi_r$ are periodic functions with frequency $\Upsilon_r$, and $\Delta t_z$ with $\Delta\phi_z$ are periodic functions with frequency $\Upsilon_z$.

It is convenient to define frequencies with respect to coordinate (Killing) time as
\begin{subequations}
\begin{align}
    \Omega_r &= \frac{\Upsilon_r}{\Gamma} \, , \\
    \Omega_z &= \frac{\Upsilon_z}{\Gamma}\, , \\
    \Omega_\phi &= \frac{\Upsilon_\phi}{\Gamma} \, ,
\end{align}
\end{subequations}
but the system is not periodic in coordinate time and these frequencies should be understood as average frequencies.

The motion is often parametrized by its orbital parameters: the semi-latus rectum $p$, the eccentricity $e$ and the inclination angle $ I$ which are defined from the turning points as
\begin{align}
    r_1 &= \frac{M p}{1-e} \; , & r_2 &= \frac{M p}{1+e} \; , & z_1 &= \sin I
\end{align}
where $0 < I < \pi/2$ for prograde orbits and $\pi/2 < I < \pi$ for retrograde orbits. Analytic expressions for the constants of motion in terms of the orbital parameters can be found in \cite{Drasco:2005kz}. Fujita and Hikida gave analytical expressions for the frequencies and coordinates in \cite{Fujita:2009}.

\section{Source term}
\label{app:source}

In this Appendix we present explicit expressions for the functions appearing in the source term for the calculation of the partial amplitudes in Eq.~\eqref{eq:Ipmlmomega}.

Whereas $A^{\rm m}_{ab}$ is entirely given by Eq.~\eqref{eq:Amab} with $P_a = \mu u_a$ and $v_a = u_a$ in the linear order, the terms in $A^{\rm d}_{ab}$ can be expressed with NP spin coefficients as
\begin{widetext}
\begin{subequations}
\begin{align}
    S^{c d} \gamma_{ndc} &= S_{ln} (\gamma + \bar{\gamma}) + S_{n\bar{m}} (-\bar{\pi} + \bar{\alpha} + \beta ) + S_{nm} (-\pi + \alpha + \bar{\beta}) + S_{m\bar{m}} (-\mu + \bar{\mu}) \; , \\
    S^{c d} \gamma_{\bar{m}dc} &= S_{ln} ( \pi + \bar{\tau} ) + S_{n\bar{m}} \bar{\rho}  + S_{nm} (\alpha + \bar{\beta}) + S_{l\bar{m}} (-\bar{\gamma} + \gamma) + S_{m\bar{m}} (-\alpha + \bar{\beta}) \; , \\
    S^{c}{}_{n} \gamma_{ndc} u^d &= S_{ln} (\gamma + \bar{\gamma}) u_n + S_{n\bar{m}} ( (\bar{\alpha} + \beta) u_n - \mu u_m ) + S_{nm} ( (\alpha + \bar{\beta} ) u_n + \bar{\mu} u_{\bar{m}} ) \; , \\
    S^{c}{}_{\bar{m}} \gamma_{\bar{m}dc} u^d &= S_{n\bar{m}} ( -\pi u_l ) + S_{l\bar{m}} ( \bar{\tau} u_n - (\bar{\gamma}-\gamma) u_{\bar{m}} ) - S_{m\bar{m}} ( - (-\alpha + \bar{\beta}) u_{\bar{m}} ) \; , \\
    S^{c}{}_{(n} \gamma_{\bar{m})dc} u^d &= ( S_{ln} ( \bar{\tau} u_n - (\bar{\gamma}-\gamma) u_{\bar{m}} ) + S_{n\bar{m}} ( \bar{\rho} u_n - \mu u_l - (\bar{\alpha}-\beta + \bar{\pi}) u_{\bar{m}} -\pi u_m ) \nonumber \\
    & \quad S_{nm} (  - (-\alpha + \bar{\beta}) u_{\bar{m}} ) + S_{l\bar{m}} (\gamma + \bar{\gamma}) u_n - S_{m\bar{m}} ( (\alpha + \bar{\beta}) u_n - \bar{\mu} u_{\bar{m}} ) )/2
\end{align}
\end{subequations}
\end{widetext}

The tetrad components of the spin tensor for $\sigma_\perp = 0$ can be expressed as
\begin{subequations}
\begin{align}
    S_{ln} &= - \sigma_{\parallel} \frac{r (\hat{K}-a^2 z^2)}{\sqrt{\hat{K}} \Sigma} \; , & S_{nm} &= \sigma_\parallel \frac{\zeta}{\sqrt{\hat{K}}} u_m u_n \; , \\
    S_{l\bar{m}} &= - \sigma_\parallel \frac{\zeta}{\sqrt{\hat{K}}} u_l u_{\bar{m}} \; , & S_{m\bar{m}} &= \sigma_{\parallel} \frac{i a z (\hat{K} + r^2)}{\sqrt{\hat{K}}\Sigma}\; , 
\end{align}
\end{subequations}
while the terms from the partial derivative for the dipole term have the form
\begin{subequations}
\begin{align}
    i (\omega S^t{}_{n} - m S^\phi{}_{n}) &= \frac{a \omega (1-z^2) - m }{\sqrt{2(1-z^2)}\Sigma} (\zeta S_{n\bar{m}} - \bar{\zeta} S_{nm}) \nonumber\\ & \quad -\frac{i K}{2\Sigma} S_{ln} \; ,  \\
    i (\omega S^t{}_{\bar{m}} - m S^\phi{}_{\bar{m}}) &= -i K \qty(\frac{S_{n\bar{m}}}{\Delta} + \frac{S_{l\bar{m}}}{2\Sigma}) \nonumber\\ & \quad + \frac{a \omega (1-z^2) - m}{\sqrt{2(1-z^2)}\zeta} S_{m\bar{m}} \; , 
\end{align}
\begin{align}
    S^r{}_n &= \frac{\Delta}{2\Sigma} S_{ln} \; , \\
    S^r{}_{\bar{m}} &= -S_{n\bar{m}} + \frac{\Delta}{2\Sigma} S_{l\bar{m}} \; , \\
    S^z{}_n &= \frac{\sqrt{1-z^2}(S_{n\bar{m}}\zeta + S_{nm}\bar{\zeta})}{\sqrt{2}\Sigma} \; , \\
    S^z{}_{\bar{m}} &= - \frac{\sqrt{1-z^2}S_{m\bar{m}}}{\sqrt{2}\zeta}\; , 
\end{align}
\end{subequations}
where $K = (r^2+a^2)\omega - a m$. The functions $f^{(i)}_{ab}$ are given by
\begin{subequations}
\begin{align}
    f^{(0)}_{nn} &= -\frac{2\zeta^2}{\Delta^2} \left( \mathcal{L}^\dag_1 \mathcal{L}^\dag_2 -2ia\zeta^{-1} \sqrt{1-z^2} \mathcal{L}^\dag_2 \right)S \, , \\
    f^{(0)}_{n\bar{m}} &= \frac{2\sqrt{2}\zeta^2}{\bar{\zeta} \Delta} \bigg( \left( \frac{i K}{\Delta} + \zeta^{-1} + \bar{\zeta}^{-1} \right) \mathcal{L}^\dag_2 \nonumber\\ &\hphantom{=} - a \sqrt{1-z^2} \frac{K}{\Delta} \left( \bar{\zeta}^{-1} - \zeta^{-1} \right) \bigg)S \, , \\
    f^{(1)}_{n\bar{m}} &= \frac{2\sqrt{2}\zeta^2}{\bar{\zeta} \Delta} \left( \mathcal{L}^\dag_2 + i a \sqrt{1-z^2} \left( \bar{\zeta}^{-1} - \zeta^{-1} \right) \right)S \, , \\
    f^{(0)}_{\bar{m}\bar{m}} &= \frac{\zeta^2}{\bar{\zeta}^2} \left( i \partial_r \left( \frac{K}{\Delta} \right) - 2 i \zeta^{-1} \frac{K}{\Delta} + \left( \frac{K}{\Delta} \right)^2 \right)S \, , \\
    f^{(1)}_{\bar{m}\bar{m}} &= - \frac{2 \zeta^2}{\bar{\zeta}^2} \left( \zeta^{-1} + i \frac{K}{\Delta} \right)S \, , \\
    f^{(2)}_{\bar{m}\bar{m}} &= - \frac{\zeta^2}{\bar{\zeta}^2} S  \, ,
\end{align}
\end{subequations}
where
\begin{equation}
    \mathcal{L}^\dag_n = -\sqrt{1-z^2}\qty( \partial_z - \frac{m - n z}{1-z^2} + a \omega ) \; .
\end{equation}

\section{Trajectory}
\label{app:trajectory}

In this Appendix we present some formulas we derived to calculate the linear in spin contribution to the trajectory. We use the tetrad from Eqs.~(47)--(51) in \cite{vandeMeent:2020} where $\Tilde{e}_2^\mu$ and $e_3^\mu$ have opposite sign to align $e_3^\mu$ with total angular momentum and to have right-handed system. Then the right hand side of MPD equations can be written as
\begin{equation}
    f^\mu_{\rm MPD} = - M e_A^{\mu} \eta^{AB} R_{B0CD} S^{CD} \, ,
\end{equation}
where $R_{B0CD}$ are components of the Riemann tensor in the Marck tetrad. Because of the way this tetrad is constructed \cite{Witzany:2019} and the fact that the Riemann tensor has a simple form in the Kinnersley tetrad, the components can be simplified to
\begin{widetext}
\begin{subequations}
\begin{align}
    R_{1012} &= \frac{3 r \sqrt{\left(\hat{K}+r^2\right) \left(\hat{K}-a^2 z^2\right)} \left(a^6 z^6-5 a^4 z^4 \left(\hat{K}+2 r^2\right)+5 a^2 r^2 z^2 \left(2 \hat{K}+r^2\right)-\hat{K} r^4\right)}{\hat{K} \Sigma^5} \; , \\
    R_{3012} &= \frac{a z \qty( -a^6 z^6 \left(\hat{K}+3 r^2\right)+a^4 z^4 \left(3 \hat{K}^2+31 \hat{K} r^2+30 r^4\right)-5 a^2 r^2 z^2 \left(3 \hat{K}+r^2\right) \left(2 \hat{K}+3 r^2\right)+3 \hat{K} r^4 \left(5 \hat{K}+3 r^2\right) )}{\hat{K} \Sigma^5} \; , \\
    R_{2013} &= \frac{a z \left(a^2 z^2-3 r^2\right)}{\Sigma^3} \; , \\
    R_{1023} &= -R_{3012}+R_{2013} \; , \\
    R_{3023} &= R_{1012} \; , 
\end{align}
\end{subequations}
\end{widetext}
and $R_{2012}=R_{1013}=R_{3013}=R_{2023}=0$.
The functions $\mathcal{R}_{t,\phi}$, $\mathcal{J}$, $\mathcal{V}$, and $\mathcal{P}$ from Eqs.~(3.24), (4.62), and (4.63) in \cite{Drummond:2022b} can be simplified to
\begin{subequations}\label{eq:traj_source}
\begin{align}
    \mathcal{R}_t &= \Sigma f_t^{\rm MPD} \; , \\
    \mathcal{R}_\phi &= \Sigma f_\phi^{\rm MPD} \; , \\
    \mathcal{J} &= - \Sigma^2 f^r_{\rm MPD} + \mathcal{I}_2 \delta u_{t}^S + \mathcal{I}_3 \delta u_\phi^S \; , \\
    \mathcal{V} &= - \Sigma^2 f^\theta_{\rm MPD} + \mathcal{U}_2 \delta u_{t}^S + \mathcal{U}_3 \delta u_\phi^S \; , \\
    \mathcal{P} &= \mathcal{N}_2 \delta u_{t}^S + \mathcal{N}_3 \delta u_\phi^S \; ,
\end{align}
\end{subequations}
where $\mathcal{I}_{2,3}$, $\mathcal{U}_{2,3}$ and $\mathcal{N}_{2,3}$ can be found in the supplemental material of \cite{Drummond:2022a}. These simplifications make the calculation of the trajectory significantly faster.

\section{Trajectories and fluxes in time domain}
\label{app:time_domain}

In this Appendix we describe our procedure to calculate trajectories and GW fluxes in the time domain in order to compare them with the frequency domain results.

\begin{figure}[tb!]
    \centering
    \includegraphics[width=0.45\textwidth]{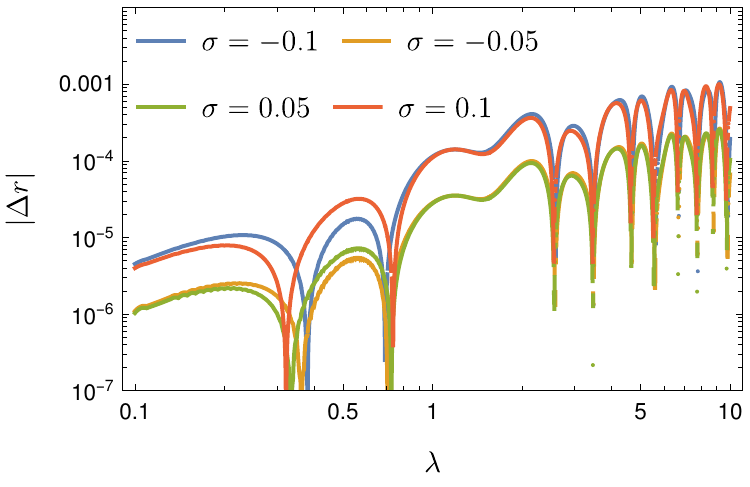}
    \caption{Difference between the time domain calculation of $r$ with the full MPD equations and linearized  in spin frequency domain calculation of $r$ for $a=0.9M$, $p=12.0$, $e=0.2$, $I = 60^\circ$ and different spins. The difference behaves as $\order{\sigma^2}$ and grows linearly in $\lambda$ on average, because of the $\order{\sigma^2}$ difference in the frequencies.}
    \label{fig:trajectory_difference}
\end{figure}

First, we calculate the orbits using the full (non\-linearized in spin) MPD equations~\eqref{eq:MPEQs} in the time domain. The initial conditions have been chosen such that the orbits are at most $\order{\sigma^2}$ from orbits with given orbital parameters in the frequency domain. As initial conditions we choose $E$, $J_z$, $r$, $\theta$, $u^r$, $s^r$ and $s^\theta$ according to the values computed in the frequency domain. Then, we find the other initial conditions from Eqs.~\eqref{eq:SSC}, \eqref{eq:mass}, \eqref{eq:spin_magnitude} and \eqref{eq:COM}. For the evolution we used an implicit Gauss-Runge-Kutta integrator which is described in \cite{Lukes-Gerakopoulos:2014}. In Fig.~\ref{fig:trajectory_difference} we plot for several spins the difference 
\begin{align*}
\Delta r = r_{\rm td}(\lambda) - \hat{r}(\Upsilon_r \lambda) - r^S(\Upsilon_r \lambda, \Upsilon_z \lambda)\, ,    
\end{align*}
where $r_{\rm td}(\lambda)$ is the evolution computed in time domain. It can be seen that the difference for $\sigma = \pm 0.1$ is four times larger than the difference for $\sigma=\pm 0.05$, thus it is indeed $\order{\sigma^2}$.

This trajectory was then used as an input to \textit{Teukode} which numerically solves the Teukolsky equation. The output is the energy flux at infinity which must be averaged to compare it with the frequency domain result. For nearly spherical orbits it is straightforward since at linear order in spin the flux has period $2\pi/\Omega_z$. Thus, we can average the flux over several periods which have been calculated using the frequency domain approach. 

For generic orbits the averaging procedure is more challenging, since the flux is not strictly periodic and it contains contributions from all the combinations of the frequencies $\Omega_r$ and $\Omega_z$. This issue was resolved by consecutive moving averages with different periods. The main contribution to the oscillations of the flux comes from the radial motion between the pericentre and apocentre. Thus, we first compute the moving average of the time series with period $2\pi/\Omega_r$ to smooth-out the data. Then, we perform several other moving averages with periods $2\pi/\Omega_z$ and combinations $2\pi/(n\Omega_r + k\Omega_z)$. After several such averages, the time series is too short for another moving average, so we average all the remaining datapoints. This procedure appears to be reliable, since the results match the frequency domain calculations.

\section{Plots and data tables}
\label{app:tables}

In this Appendix we show several plots of our frequency domain results and list the values for reference.

\begin{figure}[tb!]
    \centering
    \includegraphics[width=0.23\textwidth]{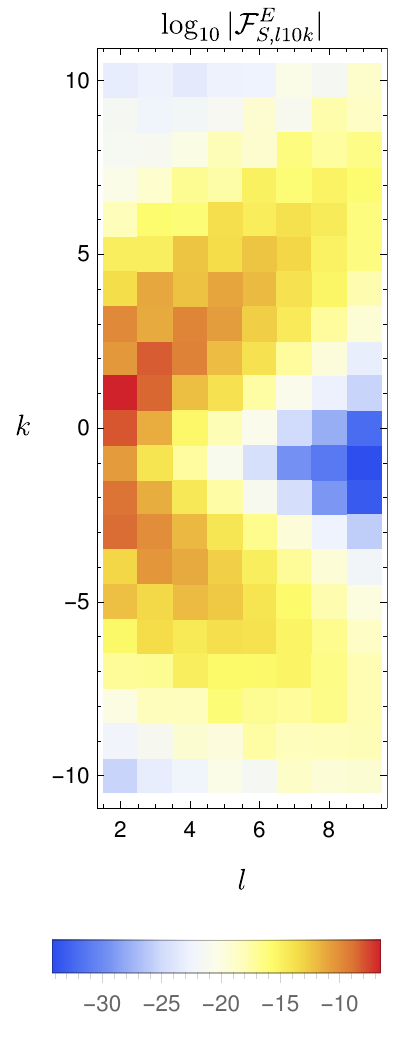}
    \includegraphics[width=0.23\textwidth]{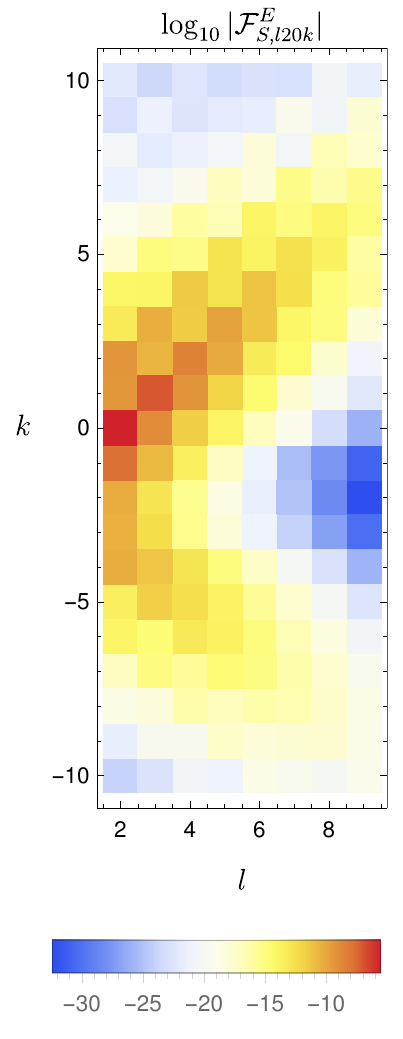}
    \includegraphics[width=0.23\textwidth]{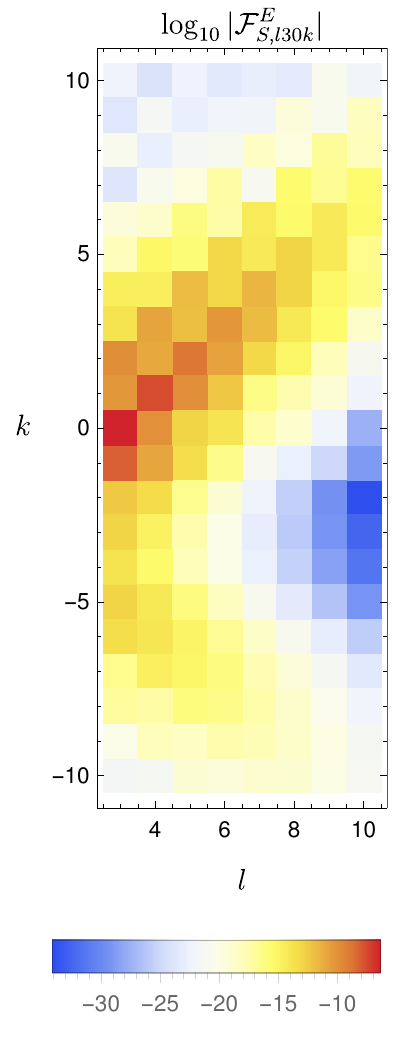}
    \caption{Linear in spin parts of the energy fluxes from nearly spherical orbits with $a=0.9M$, $p=10.0$, $I=30^\circ$ for different $l$, $k$ modes and $m=1,2,3$.}
    \label{fig:fluxes_spherical}
\end{figure}

In Fig.~\ref{fig:fluxes_spherical} we plot the linear in spin part of the total energy flux from a nearly spherical orbit for different $l$, $m$ and $k$. From these plots we can see that the linear in spin part of the flux has a global maximum at $k=l-m$ and a local maximum around $k=-l-m$. This behavior is similar to the behavior of geodesic flux that has been reported in \cite{Kerachian:2023}.

\begin{figure}[tb!]
    \centering
    \includegraphics[width=0.23\textwidth]{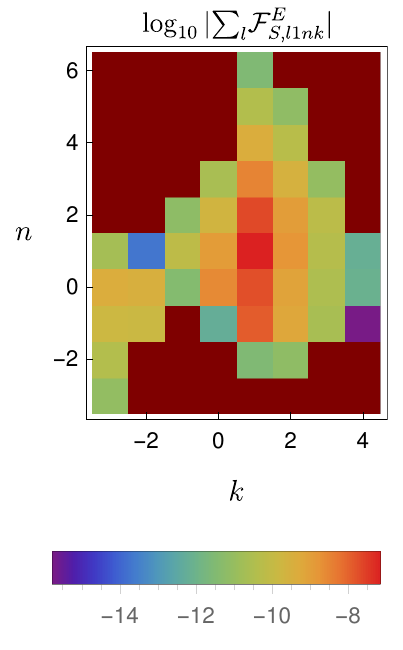}
    \includegraphics[width=0.23\textwidth]{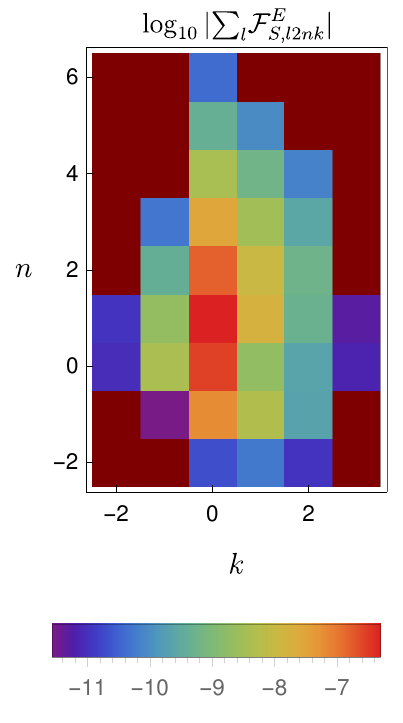}
    \includegraphics[width=0.23\textwidth]{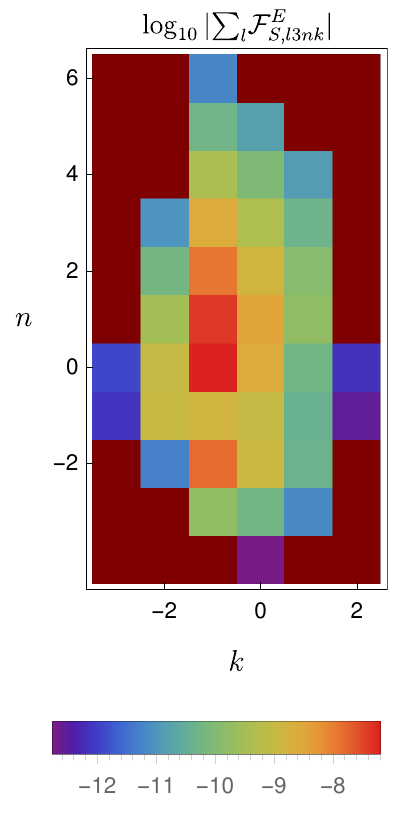}
    \caption{The linear in spin parts of the energy fluxes from generic orbits $\log_{10} \abs{\sum_l \mathcal{F}^E_{S,lmnk}}$ with $a=0.9M$, $p=12.0$, $e=0.2$, $I=30^\circ$ for different $n$, $k$ modes summed over $l$ for $m=1$ (top left), $m=2$ (top right), $m=3$ (bottom).}
    \label{fig:fluxes_generic}
\end{figure}

In Fig.~\ref{fig:fluxes_generic} we plot the $m$, $n$ and $k$ modes of the linearized in spin flux summed over $l$ for a generic orbit. Because of the computational costs, we calculated only some of the $l$, $m$, $n$, $k$ modes. We can see that the maximal mode is at $n=1$ and $k=2-m$. 

For reference, we list the $m$ modes of the linear in spin part of the energy flux for spherical orbits in Table~\ref{tab:spherical} and some of the $l$, $m$, $n$, $k$ modes from generic orbits in Table~\ref{tab:generic}.

\begin{table}[htb!]
    \centering
    \begin{tabular}{r|c|c|r}
      $I[^\circ]$ & $m$ & $\mathcal{F}^{E}_{S,m}$ & \multicolumn{1}{c}{$\mathcal{F}^{J_z}_{S,m}$} \\ \hline
        30 & 1 & $-2.642 \times 10^{-7}$ & $-2.446 \times 10^{-6}$ \\
        30 & 2 & $-2.702 \times 10^{-6}$ & $-6.431 \times 10^{-5}$ \\
        30 & 3 & $-3.921 \times 10^{-7}$ & $-1.016 \times 10^{-5}$ \\
        60 & 1 & $-1.533 \times 10^{-6}$ & $-1.891 \times 10^{-5}$ \\
        60 & 2 & $-2.177 \times 10^{-6}$ & $-5.110 \times 10^{-5}$ \\
        60 & 3 & $-2.223 \times 10^{-7}$ & $-5.463 \times 10^{-6}$ \\
        120 & 1 & $-4.175 \times 10^{-6}$ & $3.021 \times 10^{-5}$ \\
        120 & 2 & $-1.796 \times 10^{-6}$ & $3.597 \times 10^{-5}$ \\
        120 & 3 & $-1.730 \times 10^{-7}$ & $4.020 \times 10^{-6}$ \\
        150 & 1 & $-2.859 \times 10^{-6}$ & $3.280 \times 10^{-5}$ \\
        150 & 2 & $-6.930 \times 10^{-6}$ & $1.658 \times 10^{-5}$ \\
        150 & 3 & $-1.069 \times 10^{-6}$ & $2.723 \times 10^{-5}$ \\
    \end{tabular}
    \caption{Linear in spin parts of the total energy fluxes and the angular momentum fluxes from nearly spherical orbits for given inclination $I$ and azimuthal number $m$. The fluxes are summed over $l$  and $k$.}
    \label{tab:spherical}
\end{table}

%%%%%%%%%%%%%%%%%%
 \clearpage
 \onecolumngrid
%%%%%%%%%%%%%%%%%%

\begin{table}[htb!]
    \centering
    \begin{tabular}{c|c|c|c|r|r|r|r}
      $m$ & $l$ & $n$ & $k$ & \multicolumn{1}{c|}{$\Re{C^+_{S,lmnk}}$} & \multicolumn{1}{c|}{$\Im{C^+_{S,lmnk}}$} & \multicolumn{1}{c|}{$\Re{C^-_{S,lmnk}}$} & \multicolumn{1}{c}{$\Im{C^-_{S,lmnk}}$}  \\ \hline
1 & 2 & 0 & 1 & $4.8962 \times 10^{-6}$ & $-1.6020 \times 10^{-6}$ & $-5.2716 \times 10^{-6}$ & $-2.7823 \times 10^{-7}$ \\ 
1 & 2 & 1 & 1 & $9.9514 \times 10^{-6}$ & $-2.7846 \times 10^{-6}$ & $3.8592 \times 10^{-6}$ & $8.7391 \times 10^{-7}$ \\ 
1 & 2 & 2 & 1 & $7.3027 \times 10^{-6}$ & $-2.1468 \times 10^{-6}$ & $3.7407 \times 10^{-6}$ & $7.8065 \times 10^{-7}$ \\ 
1 & 2 & 3 & 1 & $3.6008 \times 10^{-6}$ & $-1.1232 \times 10^{-6}$ & $1.7742 \times 10^{-6}$ & $3.8842 \times 10^{-7}$ \\ 
1 & 3 & 0 & 2 & $-9.4587 \times 10^{-8}$ & $1.3025 \times 10^{-7}$ & $-5.8770 \times 10^{-7}$ & $-4.1971 \times 10^{-7}$ \\ 
1 & 3 & 1 & 2 & $-8.8801 \times 10^{-7}$ & $-1.7932 \times 10^{-6}$ & $-3.2098 \times 10^{-7}$ & $-2.1754 \times 10^{-7}$ \\ 
1 & 3 & 2 & 2 & $-9.8569 \times 10^{-7}$ & $-1.9468 \times 10^{-6}$ & $-1.1060 \times 10^{-7}$ & $-6.4897 \times 10^{-8}$ \\ 
1 & 3 & 3 & 2 & $-6.6574 \times 10^{-7}$ & $-1.2388 \times 10^{-6}$ & $-3.7354 \times 10^{-8}$ & $-1.4854 \times 10^{-8}$ \\ 
2 & 2 & 0 & 0 & $-1.9890 \times 10^{-5}$ & $5.8986 \times 10^{-6}$ & $-7.5636 \times 10^{-6}$ & $-3.6977 \times 10^{-6}$ \\ 
2 & 2 & 1 & 0 & $-3.6535 \times 10^{-5}$ & $1.0473 \times 10^{-5}$ & $-2.8727 \times 10^{-5}$ & $-9.5114 \times 10^{-6}$ \\ 
2 & 2 & 2 & 0 & $-2.8239 \times 10^{-5}$ & $8.6430 \times 10^{-6}$ & $-2.1354 \times 10^{-5}$ & $-7.5839 \times 10^{-6}$ \\ 
2 & 2 & 3 & 0 & $-1.5302 \times 10^{-5}$ & $4.9730 \times 10^{-6}$ & $-1.0601 \times 10^{-5}$ & $-4.1408 \times 10^{-6}$ \\ 
2 & 3 & 0 & 1 & $8.2420 \times 10^{-7}$ & $9.2288 \times 10^{-7}$ & $6.8449 \times 10^{-7}$ & $1.3381 \times 10^{-6}$ \\ 
2 & 3 & 1 & 1 & $3.8727 \times 10^{-6}$ & $7.7467 \times 10^{-6}$ & $-1.2893 \times 10^{-7}$ & $-2.4943 \times 10^{-7}$ \\ 
2 & 3 & 2 & 1 & $4.3094 \times 10^{-6}$ & $8.2104 \times 10^{-6}$ & $-4.2520 \times 10^{-7}$ & $-8.2810 \times 10^{-7}$ \\ 
2 & 3 & 3 & 1 & $3.0636 \times 10^{-6}$ & $5.4471 \times 10^{-6}$ & $-3.3219 \times 10^{-7}$ & $-6.4545 \times 10^{-7}$ \\ 
3 & 3 & 0 & 0 & $-2.2312 \times 10^{-6}$ & $-3.3582 \times 10^{-6}$ & $-3.0967 \times 10^{-7}$ & $-1.2954 \times 10^{-6}$ \\ 
3 & 3 & 1 & 0 & $-8.9746 \times 10^{-6}$ & $-1.7781 \times 10^{-5}$ & $5.0351 \times 10^{-7}$ & $3.8648 \times 10^{-6}$ \\ 
3 & 3 & 2 & 0 & $-1.0099 \times 10^{-5}$ & $-1.8845 \times 10^{-5}$ & $6.2534 \times 10^{-7}$ & $4.8041 \times 10^{-6}$ \\ 
3 & 3 & 3 & 0 & $-7.3942 \times 10^{-6}$ & $-1.2830 \times 10^{-5}$ & $4.0686 \times 10^{-7}$ & $3.3295 \times 10^{-6}$ \\ 
3 & 4 & 0 & 1 & $1.1671 \times 10^{-6}$ & $-4.1185 \times 10^{-7}$ & $1.9807 \times 10^{-7}$ & $-2.0808 \times 10^{-7}$ \\ 
3 & 4 & 1 & 1 & $-3.6720 \times 10^{-6}$ & $2.6844 \times 10^{-6}$ & $4.2934 \times 10^{-8}$ & $-4.2764 \times 10^{-8}$ \\ 
3 & 4 & 2 & 1 & $-5.6428 \times 10^{-6}$ & $4.2331 \times 10^{-6}$ & $-8.4756 \times 10^{-8}$ & $9.4155 \times 10^{-8}$ \\ 
3 & 4 & 3 & 1 & $-4.6912 \times 10^{-6}$ & $3.7655 \times 10^{-6}$ & $-1.0502 \times 10^{-7}$ & $1.1655 \times 10^{-7}$ \\ 
    \end{tabular}
    \caption{Real and imaginary parts of the linear in spin parts of amplitudes computed at infinity and at the horizon for given $l$, $m$, $n$ and $k$ of a generic orbit with $a=0.9M$, $p=12$, $e=0.2$, $I=30^\circ$.}
    \label{tab:generic}
\end{table}

%%%%%%%%%%%%%%%%%%
\twocolumngrid
%%%%%%%%%%%%%%%%%%

\bibliography{paper}

\end{document}